\documentclass[HARVARD,Times1COL]{WileyNJDv5} %STIX1COL,STIX2COL,STIXSMALL

\articletype{Research Article}%

\received{Date Month Year}
\revised{Date Month Year}
\accepted{Date Month Year}
\journal{Statistics in Medicine}
\volume{00}
\copyyear{2024}
\startpage{1}

\begin{document}

\title{Penalized GEE for Complex Carry-Over in Repeated-Measures Crossover Designs}

\author[1]{Nelson Alirio Cruz}

\author[2]{Kalliopi Mylona}

\author[3]{Oscar Orlando Melo}

\authormark{Cruz, N.A \textsc{et al.}}
\titlemark{Penalized GEE for Complex Carry-Over in Repeated-Measures Crossover Designs}

\address[1]{Profesor Visitante, Universitat de les Illes Balears, Departament de Matemàtiques i Informàtica, Phone: +34 637 54 6888, Palma de Mallorca, España}

\address[2]{Statistics Group, Department of Mathematics, Faculty of Natural, Mathematical \& Engineering Sciences, King's College London}
            
\address[3]{Profesor Titular, Departamento de Estadística, Facultad de Ciencias,
            Universidad Nacional de Colombia, Bogotá, Colombia}

\corres{Corresponding author. \email{nelson-alirio.cruz@uib.es}}

%\fundingInfo{Text}
%\JELinfo{ejlje}

\abstract[Abstract]{Crossover designs are commonly employed in clinical and behavioral research, yet the statistical models used to analyze them often rely on unrealistic assumptions—either ignoring carry-over effects or modeling them as simple and homogeneous across treatment sequences. However, carry-over effects are frequently complex, varying by treatment order and interaction, and until now, no statistical methodology had been formally established to estimate such complex effects.

This paper introduces a penalized semiparametric Generalized Estimating Equations (GEE) approach designed to estimate complex carry-over effects in crossover designs with repeated measurements. We first derive identifiability conditions under which complex carry-over effects become estimable. They then provide theoretical guarantees—building on an extension of the sandwich variance formula—showing that the proposed penalized estimator performs consistent variable selection and achieves asymptotic normality for the non-zero components.

Through simulation studies and application to real data, the methodology demonstrates improved estimation accuracy when complex carry-over effects are present, outperforming models that assume simple or no carry-over. This work represents the first rigorous and generalizable approach for modeling complex carry-over effects in repeated-measures crossover designs.}

\keywords{carry-over effect, Generalized estimating equations, Kronecker correlation, Parameter estimability, Correlated data}

\maketitle

\renewcommand\thefootnote{\fnsymbol{footnote}}
\setcounter{footnote}{1}
\section{Introduction}
Crossover experimental designs apply a sequence of treatments to the same experimental units over multiple periods, facilitating efficient comparisons under controlled conditions. A major challenge in analyzing such designs is accounting for carry-over effects—the residual influence of a treatment persisting into subsequent periods. These effects may be classified by order, where first-order affects only the immediately following treatment, and second-order impacts treatments two periods later. Furthermore, carry-over effects can be simple, exerting a uniform influence on all subsequent treatments, or complex, where the magnitude and direction of the effect depend on specific treatment sequences \citep{vegas2016crossover}.

Classical studies \citep{fleiss1989, senn1992simple} have shown that ignoring complex carry-over effects or simplifying them to first-order effects may induce bias in treatment effect estimates. Interestingly, assuming simple carry-over effects tends to produce less bias than ignoring carry-over altogether, but the common scenario of complex carry-over effects remains largely unaddressed due to identifiability challenges—especially in crossover designs with only one observation per experimental unit per period.

A recent review of the literature reveals that most crossover studies either assume simple carry-over effects or disregard them completely, even when repeated measurements per unit and period are available \citep{basu2010joint, josephy2015within, hao2015explicit, lui2015test, rosenkranz2015analysis, grayling2018blinded, madeyski2018effect, kitchenham2018corrections, oh2003bayesian, curtin2017meta, li2018power, jankar2020optimal, cruz2023correlation}. This gap points to the need for methodologies capable of reliably estimating complex carry-over effects in longitudinal crossover data.

A notable applied example is the work environment study by \cite{pitchforth2020work}, which implemented a Williams design with 288 participants allocated to four groups experiencing four office layouts (A: Menu, B: Control, C: Nest, D: Campus). Each period lasted two weeks, and occupancy was repeatedly recorded every 5 minutes through infrared sensors, yielding 96 binary observations per experimental unit and period (see Table \ref{tableZA}). Prior analyses assumed simple carry-over effects \citep{jankaroptimal}.

\begin{table}[ht]
    \centering
    \begin{tabular}{c|c|c|c|c|}
        Sequence & Period 1 & Period 2 & Period 3 & Period 4 \\
        \hline
        Group 1 & B & A & D & C \\
        Group 2 & C & D & A & B \\
        Group 3 & D & B & C & A \\
        Group 4 & A & C & B & D \\
        \hline
    \end{tabular}
    \caption{Structure of the crossover design in the work environment experiment \citep{pitchforth2020work}.}
    \label{tableZA}
\end{table}

Until recently, estimation of complex carry-over effects remained elusive. \citealp{cruz2023correlation} proposed a semiparametric Generalized Estimating Equations (GEE) approach to estimate simple carry-over effects when multiple observations per unit and period exist. However, conditions for identifiability and estimation of complex carry-over effects have not been thoroughly examined, nor have these methods been assessed in common crossover designs with repeated measurements. Additionally, the challenge of estimating a large number of carry-over parameters under complex scenarios has not been addressed.

Recent advances in penalized GEE methods for high-dimensional or irregular longitudinal data \citep{wang2012penalized, lee2022doubly, lu2024penalized, jaman2025penalized}—implemented in packages such as \texttt{LassoGEE} \citep{LasoGEE} and \texttt{PGEE} \citep{PGEE}—offer promising tools but currently lack reproducible implementations tailored to penalized GEE for complex carry-over effects in crossover designs with repeated measurements.

In this paper, we first formalize the conditions under which complex carry-over effects are identifiable in crossover designs with repeated measures. Next, we introduce a semiparametric penalized GEE framework incorporating spline-based regularization to flexibly estimate both time and carry-over effects across a broad class of response distributions. We conduct simulation studies to evaluate the methodology under various scenarios, including AB/BA designs, and finally apply the approach to real data from a work environment experiment. Our results demonstrate superior performance relative to models assuming simple or no carry-over effects, and establish practical guidelines requiring at least five within-period observations per experimental unit for stable and reliable estimation.

\section{Crossover designs with repeated measures}
In a crossover design with $S$ sequences of length $P$, let $L$ be the number of repeated observations in the $i$-th experimental unit during the $j$-th period. The corresponding response vector is defined as
\begin{equation}\label{observation}
    \pmb{Y}_{ij} = \left(Y_{ij1}, \ldots, Y_{ijL} \right)^\top.
\end{equation}
The complete observation vector for the $i$-th experimental unit across all periods is denoted by
\begin{equation}
    \pmb{Y}_i = \left(\pmb{Y}_{i1}, \ldots, \pmb{Y}_{iP} \right)^\top,
\end{equation}
which has dimension $PL \times 1$, with $i = 1, \ldots, n$ indexing the units per sequence.

Each individual response $Y_{ijk}$ is assumed to follow the semiparametric additive model:
\begin{align}
    E(Y_{ijk}) &= \mu_{ijk}, \nonumber \\
    g(\mu_{ijk}) &= \pmb{x}_{ijk}^\top \pmb{\beta} + f({Z}_{ijk}) + \sum_{c=1}^{C} f_c({Z}_{ijk}), \label{ectruth}
\end{align}
where $g(\cdot)$ is a known link function, $\pmb{x}_{ijk}$ is the covariate vector associated with the $k$-th measurement of subject $i$ in period $j$, and $\pmb{\beta}$ represents the parametric coefficients. ${Z}_{ijk}$ is the time of medition of $Y_{ijk}$ response, the function $f({Z}_{ijk})$ captures the nonlinear trend associated with the measurement time ${Z}_{ijk}$, and each $f_c({Z}_{ijk})$ models a specific carry-over effect resulting from the treatment administered in the preceding period. As carry-over effects are not present in the first period, all $f_c({Z}_{i1k})$ are set to zero.

The total number of complex carry-over functions, denoted by $C$, depends on the number of possible ordered treatment pairs. For instance, in a crossover design involving four treatments (A, B, C, D) arranged across four periods as in Table~\ref{tableZA}, there are $4 \times 3 = 12$ first-order complex carry-over functions. Each function corresponds to a directed treatment pair $(T_p \rightarrow T_q)$, reflecting the effect of treatment $T_p$ in period $j-1$ on treatment $T_q$ in period $j$. Specifically, the twelve functions are: the effect of A on B, C, and D; the effect of B on A, C, and D; the effect of C on A, B, and D; and the effect of D on A, B, and C. These effects are modeled independently, allowing for treatment-specific and time-varying dynamics in carry-over behavior.

The identifiability and estimation of complex carry-over effects are nontrivial and depend critically on both the structure of the design and the number of observations within each period. The following section formalizes the sufficient conditions under which these effects can be identified and consistently estimated within a penalized GEE framework.

\section{Estimability conditions}
The identifiability and estimation of the complex carry-over effects depend on the experimental design and data structure. The following assumptions establish sufficient conditions for the estimability of the treatment, period, and carry-over effects in the penalized GEE framework.

\begin{assumption}\label{as1}
 Each treatment must appear in at least two different sequences.
\end{assumption}
\begin{assumption}\label{as2}
Each treatment must appear in more than one period, and each period must have more than one treatment.
\end{assumption}
\begin{assumption}\label{as3}
Each carry-over effect included in the model must appear in at least one observation in the design; that is, the corresponding indicator must be non-zero for at least one unit-period.
\end{assumption}
\begin{assumption}\label{as4}
For each carry-over effect, the number of observations where that carry-over effect is active (i.e., the corresponding carry-over indicator is nonzero) is at least $d$.
\end{assumption}
To establish the rank properties of the design matrices involved, the following lemmas are introduced. For a given matrix $\pmb{A}$, we denote by $\mathcal{C}(\pmb{A})$ its column space, that is, the linear span of the columns of $\pmb{A}$. For any $n \in \mathbb{N}$, let $\mathbf{1}_n=(1,1,\ldots,1)^\top$ denote the vector of ones of size $n$.

\begin{lemma}\label{lema1}
Let $\pmb{A}$ be a matrix of dimension $n\times p$, $n> p$ and $\mathcal{C}(\pmb{A})$ the generated columns space by $\pmb{A}$. If $\mathbf{1}_{n} \notin \mathcal{C}(\pmb{A})$ then:
\begin{enumerate}
    \item  $\mathbf{1}_{nL} \notin \mathcal{C}(\pmb{A}\otimes \mathbf{1}_L)$.
     \item  $\mathbf{1}_{nL} \notin \mathcal{C}(\mathbf{1}_L\otimes \pmb{A})$.
\end{enumerate}
where $\mathbf{1}_{L}=(1,1,\ldots,1)^\top$ a vector of size $L$ and $\mathbf{1}_{nL}=\mathbf{1}_L\otimes\mathbf{1}_{n}$
\end{lemma}
\begin{proof}
See Appendix \ref{prooflema1}
\end{proof}
\begin{lemma}\label{lema2}
Let $\pmb{A}$ a matrix of dimension $n\times p$, $n> p$ and $rank(\pmb{A})=p$,  $\pmb{B}$ be a matrix of dimension $n\times q$, $n> q$ and $rank(\pmb{B})=q$. If $rank([A\,\vdots B])=p+q$ then $ rank([\pmb{A}\otimes \mathbf{1}_L\,\vdots\, \pmb{B}\otimes \mathbf{1}_L])=p+q$ 
\end{lemma}
\begin{proof}
See Appendix \ref{prooflema2}
\end{proof}

\begin{lemma}\label{lema3}
    Let $\pmb{A}$ be a matrix of dimension $L\times p$, $L> p$ and $rank(\pmb{A})=p$,  $\pmb{B}$ be a matrix of dimension $n\times q$, $n> q$, $rank(\pmb{B})=q$, $\pmb{B}=\{b_{ij}\}_{n\times q}$ and $b_{ij} \in \{0,1\}$, and $\pmb{D}$ be a matrix of dimension $n\times r$, $n> r$, $rank(\pmb{D})=r$, $\pmb{D}=\{d_{ij}\}_{n\times r}$ and $d_{ij} \in \{0,1\}$.  If $\mathbf{1}_{L} \notin \mathcal{C}(\pmb{A})$, then $rank([\pmb{D}\otimes\mathbf{1}_{L}\,\vdots \pmb{B} \otimes \pmb{A}]) =r+pq$.
\end{lemma}
\begin{proof}
See Appendix \ref{prooflema3}
\end{proof}

Let $\mathcal{H}$ be a Hilbert space of functions defined on the time interval $\mathcal{T} \subset \mathbb{R}$, equipped with the inner product $\langle \cdot, \cdot \rangle$.

Consider an orthonormal basis $\{\phi_1, \phi_2, \ldots, \phi_d\}$ of $\mathcal{H}$, where $d$ may be finite or infinite, which allows representing any function $f \in \mathcal{H}$ as
\[
f(t) = \sum_{l=1}^d \theta_l \phi_l(t), \quad t \in \mathcal{T},
\]
with coefficients $\theta_l \in \mathbb{R}$.

For each experimental unit $i = 1, \ldots, n$, period $j = 1, \ldots, P$, and repeated measure $k = 1, \ldots, L$ taken at times $\{Z_{ijk}\}_{k=1}^L \subset \mathcal{T}$, define the basis evaluation matrix
\[
\Phi_{ij} = \begin{bmatrix}
\phi_1(Z_{ij1}) & \phi_2(Z_{ij1}) & \cdots & \phi_d(Z_{ij1}) \\
\phi_1(Z_{ij2}) & \phi_2(Z_{ij2}) & \cdots & \phi_d(Z_{ij2}) \\
\vdots & \vdots & \ddots & \vdots \\
\phi_1(Z_{ijL}) & \phi_2(Z_{ijL}) & \cdots & \phi_d(Z_{ijL})
\end{bmatrix}_{L \times d}.
\]

This matrix $\Phi_{ij}$ evaluates the functional basis at the observed time points for experimental unit $i$ in period $j$, and is used to represent the functional time and carry-over effects in the model.

The matrix $\pmb{C}$ is a binary indicator matrix of dimension $nP \times C$, where $C$ denotes the number of first-order complex carry-over effects considered. Each column encodes the presence or absence of a specific carry-over effect for each experimental unit and period.

Let the vector $\pmb{Y}=\{\pmb{Y}_{1}^\top,\cdots,\pmb{Y}_{i}^\top, \cdots, \pmb{Y}_{n }^\top \}^\top$, where the experimental units are organized by sequences and by periods, then the design matrix associated with the model defined in the equation \eqref{ectruth} has the form:
\begin{align}\label{xxx}
     \pmb{X}=\left\{\mathbf{1}_{nPL}\, \vdots \, \pmb{T} \otimes \mathbf{1}_{L}\, \vdots \ , \pmb{\mathcal{P}} \otimes \mathbf{1}_{L}\, \vdots
     \, \mathbf{1}_{nP}\otimes \pmb{\Phi} \, \vdots \, \mathbf{C}\otimes \pmb{\Phi} \right\}_{ nPL\times q}
\end{align}
where $\mathbf{1}_{L}$ is a vector of ones of size $L$, $\pmb{T}$ is a matrix that indicates the effect of each treatment on each $i$ in period $j$, after applying estimability conditions, that is, $\sum _{d=1}^D \tau_d=0$; therefore, it is of size $nP\times (D-1)$ with $D$ number of treatments. Additionally, $\pmb{\mathcal{P}}$ is a matrix that indicates the period effect on each $i$ in the period $j$, it is of size $nP\times (P -1)$, $\pmb{\mathcal{A}}$ is the matrix of size $m+3$ associated with the solution of the B-splines equations of the third degree with $m$ nodes for the $L$ times $Z_{ijk}$ of the individual $i$ in the period $j$, it is worth clarifying that the splines are fitted without intercept, and $\pmb{\mathcal{C}}$ is a matrix which indicates the first-order complex carry-over effect on each $i$ in period $j$.

To exemplify this notation, an AB/BA crossover design with $L=10$ observations from each experimental unit within each period, and two experimental units per sequence, as seen in Table \ref{tabla11A} will be used.

\begin{table}[h!]
\caption{$2\times 2$ usual crossover design}
\label{tabla11A}
 \centering
 \begin{tabular}{|c| c c |} 
 \hline
 Sequence & Period 1 & Period 2\\ \hline
AB & 10 observations & 10 observations  \\
BA & 10 observations & 10 observations  \\
 \hline
 \end{tabular}
\end{table}
In this case, $P=2$, $D=2$, and $n=2$. In addition, $\pmb{Y}_1$ and $\pmb{Y}_2$ belong to the first sequence (AB) and $ \pmb{Y}_3$ and $\pmb{Y}_4$ belong to the second sequence (BA). It is also defined: $\tau_A=0$ as the effect of the treatment A, and $\pi_1=0$ as the effect of the period 1. Therefore,
\begin{align*}
    &\pmb{Y}_{40\times 1}  = \begin{bmatrix}
        \pmb{Y}_1\\
        \pmb{Y}_2\\
        \pmb{Y}_3\\
        \pmb{Y}_4
    \end{bmatrix} = \begin{bmatrix}
         \begin{bmatrix}
        \pmb{Y}_{11}\\
        \pmb{Y}_{12}\\
    \end{bmatrix}\\
        \begin{bmatrix}
        \pmb{Y}_{21}\\
        \pmb{Y}_{22}\\
    \end{bmatrix}\\
        \begin{bmatrix}
        \pmb{Y}_{31}\\
        \pmb{Y}_{32}\\
    \end{bmatrix}\\
       \begin{bmatrix}
        \pmb{Y}_{41}\\
        \pmb{Y}_{42}\\
    \end{bmatrix}
    \end{bmatrix}, \;
 \pmb{T} = \begin{bmatrix}
        0\\
        1\\
        0 \\
        1 \\
        1\\
        0\\
        1\\
        0
    \end{bmatrix}, \; \pmb{\mathcal{P}} = \begin{bmatrix}
        0\\
        1\\
        0\\
        1\\
        0\\
        1\\
        0\\
        1\\      
    \end{bmatrix}, \;  \pmb{C} = \begin{bmatrix}
        0 & 0\\
        1 & 0\\
        0 & 0\\
        1 & 0\\
        0 & 0\\
        0 & 1\\
        0 & 0\\
        0 & 1\\      
    \end{bmatrix}
\end{align*}
In the previous matrices, $\pmb{T}$ is equal to 1 when treatment B is present, that is, in period 2 of experimental units 1 and 2, and period 1 of experimental units 3, and 4, and the other scenarios it is equal to 0. Analogously, $\pmb{\mathcal{P}}$ is equal to 1 in period 2 and 0 in period 1. As to the matrix $\pmb{C}$, the first column is the first-order carry-over effect of A on B, that is, it is 0 if the observation is in period 1, and it is  1 in period 2 and received in period 1 treatment A, and column 2 is analogous but with the carry-over effect of B on A.

\begin{assumption}\label{as5}
The functional design matrix
\[
\pmb{\Phi}_c = \left[\phi_\ell(Z_{ijk})\right]_{\substack{1 \leq k \leq L \\ 1 \leq \ell \leq d}} \in \mathbb{R}^{L \times d},
\]
with $L \geq d$ and active measurement times $t_m$, has full column rank $d$, i.e., 
\[
\mathrm{rank}(\pmb{\Phi}_c) = d.
\]
\end{assumption}

With these lemmas, the following result on the methodology is obtained for estimating complex carry-over effects in crossover designs:
\begin{theorem}\label{te1}
Consider a crossover design with $S$ sequences and $P$ periods, where for each experimental unit $i$ and period $j$ there are $L$ repeated measurements at times $\{Z_{ijk}\}_{k=1}^L$. Suppose the time effects, and the complex carry-over effects are modeled as functions
\begin{align}
   f(Z_{ijk}) &= \sum_{\ell=1}^d \theta_{\ell} \phi_\ell(Z_{ijk}),\label{time}\\
    f_c(Z_{ijk}) &= \sum_{\ell=1}^d \theta_{c\ell} \phi_\ell(Z_{ijk}),\label{ccarry}
\end{align}
where $\{\phi_\ell\}_{\ell=1}^d$ is a known finite-dimensional functional basis (e.g., B-splines or Fourier polynomials) of dimension $d$, $\pmb{\theta} = (\theta_{1}, \ldots, \theta_{d})^\top$, and $\pmb{\theta}_c = (\theta_{c1}, \ldots, \theta_{cd})^\top$ are the coefficients to be estimated.

Then, under assumptions \ref{as1}  through \ref{as5}, the carry-over effects $f_c$ are estimable.
Consequently, the full design matrix has full column rank, including treatment effects, period effects, and carry-over effects modeled via the chosen functional basis. This ensures the unique estimability of the coefficient vector $\pmb{\theta}_c$ through linear or semiparametric estimation methods.
\end{theorem}
\begin{proof}
See Appendix \ref{prooflema4}
\end{proof}
\begin{corollary}
Assume that $f_c$ belongs to the space 
\begin{align*}
    \mathcal{H}_F &= \mathrm{span}\Bigg\{\sin\left(\frac{2\pi t}{T}\right), \cos\left(\frac{2\pi t}{T}\right),\ldots,\sin\left(\frac{K\pi t}{T}\right), \cos\left(\frac{K\pi t}{T}\right)\Bigg\}
\end{align*}
of dimension $d = 2K + 1$ for some integer $K\geq 1$. Then the carry-over effect $f_c$ is estimable if and only if the number of distinct times $t_{ijk}$ at which $f_c$ is active is at least $d$.
\end{corollary}
\begin{corollary}
Assume that $f_c$ belongs to the space $\mathcal{H}_B$ generated by a B-spline basis of order $k$ (degree $k-1$), with $m$ internal knots on $[0,T]$ excluding intercept, then, $\dim(\mathcal{H}_B) = m + k - 1$. The carry-over effect $f_c$ is estimable if and only if:
\begin{enumerate}
    \item The number of observations at which $f_c$ is active is at least $d = m + k - 1$.
    \item The matrix of B-spline evaluations at the corresponding times has full column rank.
\end{enumerate}
In particular, if $L$ repeated measurements per period satisfy $L > m + k - 1$, and these measurements are not concentrated at a few points, estimability holds.
\end{corollary}
In summary, Assumptions~\ref{as1}--\ref{as5} together with Lemmas~\ref{lema1}--\ref{lema3} establish the structural requirements under which the design matrix is of full column rank. Theorem~\ref{te1} guarantees that the treatment effects, period effects, and complex carry-over effects modeled via a functional basis are all identifiable and uniquely estimable. This theoretical framework ensures that the semiparametric estimation problem is well-posed and that the parameter vector admits a unique solution in the space spanned by the chosen basis functions. 

The next section describes the estimation methodology under these conditions, focusing on the penalized GEE approach that enables consistent and efficient estimation of the time-varying and carry-over effects while accounting for within-unit correlation structures.
\section{Estimation procedure under General Functional Bases}
The estimation approach developed in \cite{CruzSemi2023} provides a semiparametric framework for repeated measures data in crossover designs, where the response variable $Y_{ijk}$ follows a distribution in the exponential family. In that work, GEE were derived for both parametric and nonparametric components, using B-spline basis functions to model time and complex carry-over effects.

In this section, we extend that methodology by allowing the functional effects to be represented through general basis functions spanning a separable Hilbert space. This generalization enables the use of Fourier bases, wavelets, or other smooth basis systems, beyond B-splines, thereby increasing the flexibility of the model and the potential applicability to diverse temporal structures.
Let $\{\phi_1(t), \ldots, \phi_d(t)\}$ be a set of basis functions generating a separable Hilbert space $\mathcal{H}$, where each function $\phi_\ell(t)$ is square integrable over the domain of time observations $T \subset \mathbb{R}$. This basis can correspond to B-splines, Fourier series, wavelets, or any other orthonormal system defined over $T$.

The semiparametric model remains as in Equation \eqref{ectruth}, and additionally,
\begin{align}
    \mathrm{Var}(Y_{ijk}) &= \phi V(\mu_{ijk}) \nonumber \\
    \pmb{V}(\pmb{\mu}_i) &= \left[\pmb{D}(V(\mu_{ijk})^{\frac{1}{2}}) \pmb{R}(\pmb{\alpha}) \pmb{D}(V(\mu_{ijk})^{\frac{1}{2}})\right]_{PL \times PL} \label{ectruth1}
\end{align}
where $f$ and $f_c$ are assumed to lie in $\mathcal{H}$ as in Equations \eqref{time} and \eqref{ccarry}, respectively. The dimension $d$ is chosen to satisfy the smoothness–bias trade-off, subject to conditions ensuring $d/n \to 0$ as $n \to \infty$ and the condition of estimability defined in Theorem \ref{te1}. $V(\mu_{ijk})$ is the variance function of the exponential family and $R(\pmb{\alpha})$ is the associated correlation matrix.

Accordingly, the GEE for $\pmb{\theta}$, $\pmb{\theta}_c$, $\pmb{\beta}$, and the correlation parameters $\pmb{\alpha}$ become:
    \begin{align}
    U_1(\pmb{\theta}|\pmb{\beta}, \pmb{\theta}_c, \pmb{\alpha})& = \sum_{i=1}^n \left\{ \mathrm{diag}\left(\frac{\partial \mu_{ijk}}{\partial \pmb{\theta}}\right) \right\}_i \pmb{V}_{1i}^{-1} \left(\pmb{y}_i - \pmb{\mu}_i[\pmb{\beta}, \pmb{\theta}, \pmb{\theta}_c] \right)\label{u1gee}\\
    \pmb{V}_{1i}&=\left\{diag\left(V \left(\pmb{\mu}_i\left[\pmb{X}_i \pmb{\beta},\sum_{\ell=1}^d \theta_{\ell} \phi_\ell(t) ,\hat{f}_c(\pmb{Z}_{2i}) \right]  \right) \right)\right\}_i^\frac{1}{2} \times \pmb{R}(\mathbf{\pmb{ \alpha}})\times\nonumber\\ 
    & \left\{diag\left(V \left(\pmb{\mu}_i\left[\pmb{X}_i \pmb{\beta},\sum_{\ell=1}^d \theta_{\ell} \phi_\ell(t)  ,\hat{f}_c(\pmb{Z}_{2i}) \right]  \right) \right)\right\}_i^\frac{1}{2} \nonumber\\    
     U_2(\pmb{\theta}_c|\pmb{\beta}, \pmb{\theta}, \pmb{\alpha})& = \sum_{i=1}^n \left\{ \mathrm{diag}\left(\frac{\partial \mu_{ijk}}{\partial \pmb{\theta}_c}\right) \right\}_i \pmb{V}_{2i}^{-1} \left(\pmb{y}_i - \pmb{\mu}_i[\pmb{\beta}, \pmb{\theta}, \pmb{\theta}_c] \right)\label{u2gee}\\
     U_3(\pmb{\beta}|\pmb{\theta}, \pmb{\theta}_c, \pmb{\alpha}) &= \sum_{i=1}^n \left\{ \mathrm{diag}\left(\frac{\partial \mu_{ijk}}{\partial \pmb{\beta}} \right) \right\}_i \pmb{V}_{3i}^{-1} \left(\pmb{y}_i - \pmb{\mu}_i[\pmb{\beta}, \pmb{\theta}, \pmb{\theta}_c] \right)\label{u3gee}\\
     U_4(\pmb{ \alpha}|\pmb{\beta}, , \pmb{\theta}, \pmb{\theta}_c)&=\sum_{i=1}^n\left(\frac{\partial \pmb{\varepsilon}_{ik}}{\partial \pmb{ \alpha}} \right)^\top \pmb{F}_{ik}^{-1} \left(\pmb{W}_{ik} - \pmb{\varepsilon}_{ik} \right)\label{u4gee}
    \end{align}
    where $\frac{\partial \mu_{ijk}}{\partial \pmb{\theta}} = \frac{\partial \mu_{ijk}}{\partial \eta_{ijk}} \cdot \pmb{s}(Z_{1ijk})$, and similarly for other derivatives.
The next iterative process applies:
\begin{enumerate}
    \item Initialize $\pmb{\beta}^{(0)}$, $\pmb{\theta}^{(0)}$, $\pmb{\theta}_c^{(0)}$, and $\pmb{\alpha}^{(0)}$.
    \item Update $\pmb{\theta}^{(m+1)}$ from $U_1 = 0$.
    \item Update $\pmb{\theta}_c^{(m+1)}$ from $U_2 = 0$.
    \item Update $\pmb{\beta}^{(m+1)}$ from $U_3 = 0$.
    \item Update $\pmb{\alpha}^{(m+1)}$ from $U_4 = 0$.
    \item Iterate steps 2. to 5. until convergence.
\end{enumerate}
The theoretical consistency and asymptotic normality of $\hat{\pmb{\beta}}$ hold under mild assumptions on the basis functions and their approximation properties, extending Theorem 1 of \cite{cruz2023correlation} to any basis with bounded $r$-th derivative and $d = o(n)$.
When using a general functional basis, especially with large $d$, the estimation of the coefficient vectors $\pmb{\theta}$ and $\pmb{\theta}_c$ may lead to overfitting and instability due to high dimensionality. To address this, we incorporate penalization into the GEE system. This is equivalent to maximizing a penalized quasi-likelihood or, equivalently, solving modified estimating equations of the form:
\begin{align}
U_1^{\text{pen}}(\pmb{\theta}) &= U_1(\pmb{\theta}) - \lambda \mathbf{P}_1 \pmb{\theta} \label{penU1}\\
U_2^{\text{pen}}(\pmb{\theta}_c) &= U_2(\pmb{\theta}_c) - \lambda_c \mathbf{P}_2 \pmb{\theta}_c, \quad c=1, \ldots, C \label{penU2}
\end{align}
where $U_1(\pmb{\theta})$ and $U_2(\pmb{\theta}_c)$ are defined as in Equations \eqref{u1gee} and \eqref{u2gee}, respectively. The matrices $\mathbf{P}_1$ and $\mathbf{P}_2$ are symmetric positive semi-definite penalty matrices (e.g., identity for ridge penalty or discrete difference matrices for smoothness), and $\lambda, \lambda_c > 0$, $c=1, \ldots, C$ are smoothing parameters controlling the strength of the regularization.
The estimating equations for the fixed effects $\pmb{\beta}$ and the correlation parameters $\pmb{\alpha}$ remain unchanged, so:
\begin{align}
U_3(\pmb{\beta}) &= \sum_{i=1}^n \left\{ \mathrm{diag}\left(\frac{\partial \mu_{ijk}}{\partial \pmb{\beta}} \right) \right\}_i \pmb{V}_{3i}^{-1} \left(\pmb{y}_i - \pmb{\mu}_i[\pmb{\beta}, \pmb{\theta}, \pmb{\theta}_c] \right)\label{u3gee_pen}\\
U_4(\pmb{\alpha}) &=\sum_{i=1}^n\left(\frac{\partial \pmb{\varepsilon}_{ik}}{\partial \pmb{ \alpha}} \right)^\top \pmb{F}_{ik}^{-1} \left(\pmb{W}_{ik} - \pmb{\varepsilon}_{ik} \right)\label{u4gee_pen}
\end{align}
where $\pmb{F}_{ik}=\pmb{D}(V(r_{ijk}))_{q\times q}$ is a diagonal matrix, $\pmb{\varepsilon}_{ik}=E(\pmb{W}_{ik})_{q\times 1}$ and 
$\pmb{W}_{ik}= (r_{i1k}r_{i2k},$ $r_{i1k}r_{i3k},\ldots,r_{i(T-1)k}r_{iTk})^\top_{q\times 1}$,  $r_{ijk}$ is the $ijk$-th  Pearson residual and  $q={T \choose 2}$. Thus, the iterative procedure becomes:
\begin{enumerate}
    \item Initialize $\pmb{\beta}^{(0)}$, $\pmb{\theta}^{(0)}$, $\pmb{\theta}_c^{(0)}$, and $\pmb{\alpha}^{(0)}$.
    \item Update $\pmb{\theta}^{(m+1)}$ by solving $U_1^{\text{pen}} = 0$.
    \item Update $\pmb{\theta}_c^{(m+1)}$ by solving $U_2^{\text{pen}} = 0$.
    \item Update $\pmb{\beta}^{(m+1)}$ from $U_3 = 0$.
    \item Update $\pmb{\alpha}^{(m+1)}$ from $U_4 = 0$.
    \item Iterate the steps 2. to 5. until convergence.
\end{enumerate}
The inclusion of penalization does not alter the identifiability conditions described in Theorem \ref{te1}. The theorem ensures that, under appropriate assumptions on the design and basis structure, the true functional effects are estimable as long as $d/n \to 0$. The penalty terms act only to regularize the solution within the identified parameter space and do not induce confounding or identifiability issues. In fact, in high-dimensional settings, penalization improves numerical stability and convergence, and may help reduce variance without introducing asymptotic bias under proper tuning.

The inclusion of roughness penalties in the estimating equations, through regularization terms involving penalty matrices $\mathbf{P}_1$ and $\mathbf{P}_2$, introduces tuning parameters $\lambda$ and $\lambda_c$, $ c=1, \ldots, C$ that control the trade-off between bias and variance in the estimation of the functional effects $f(t)$ and $f_c(t)$. These penalization parameters play a critical role in ensuring stable and interpretable estimates, especially when the dimension $d$ of the basis expansion is large relative to the sample size or when the number of carry-over functions increases with the number of treatments.

The optimal values of $\lambda$ and $\lambda_c$,  $ c=1, \ldots, C$ may be selected using the following options:
\begin{enumerate}
    \item Quasi-likelihood-based model selection criteria adapted to the GEE framework: A commonly used approach is to minimize the \emph{Quasi-likelihood under the Independence model Criterion} (QIC), an extension of the Akaike Information Criterion (AIC) to the quasi-likelihood setting. Given a fitted penalized GEE model, the QIC is defined as
\begin{equation}
    \text{QIC}(\lambda, \lambda_1, \ldots, \lambda_C) = -2Q(\widehat{\pmb{\theta}}, \widehat{\pmb{\theta}}_c, \widehat{\pmb{\beta}}) + 2\, \mathrm{tr}(\widehat{\pmb{V}}^{-1} \widehat{\pmb{W}})
\end{equation}
where $Q(\cdot)$ is the quasi-likelihood evaluated at the fitted values, $\widehat{\pmb{V}}$ is the model-based variance, and $\widehat{\pmb{W}}$ is the robust sandwich estimator. The pair $(\lambda, \lambda_1, \ldots, \lambda_C)$ that minimizes QIC over a grid of candidate values is selected.
\item Cross-validation: procedures can be used. In the longitudinal setting, the \emph{leave-one-cluster-out} cross-validation (LOCO-CV) is often recommended. For each candidate pair $(\lambda, \lambda_1, \ldots, \lambda_C)$, the following error is computed:
\begin{equation}
    \text{CV}(\lambda, \lambda_1, \ldots, \lambda_C) = \sum_{i=1}^n \| \pmb{y}_i - \widehat{\pmb{\mu}}_i^{(-i)} \|^2
\end{equation}
where $\widehat{\pmb{\mu}}_i^{(-i)}$ is the predicted response for experimental unit $i$ when the model is trained excluding all observations from cluster $i$.
\item Generalized cross-validation: When the model admits a linear representation for the penalized components, the \emph{generalized cross-validation} (GCV) score may be used as a computationally efficient surrogate:
\begin{equation}
    \text{GCV}(\lambda, \lambda_1, \ldots, \lambda_C) = \frac{\| \pmb{y} - \widehat{\pmb{\mu}} \|^2}{\left[1 - \frac{\mathrm{trace}(\pmb{S}_{\lambda, \lambda_1, \ldots, \lambda_C})}{n} \right]^2}
\end{equation}
where $\pmb{S}_{\lambda, \lambda_1, \ldots, \lambda_C}$ is the effective smoothing matrix corresponding to the penalized components. Though challenging in a full GEE setting, this criterion can be approximated in two-step procedures where functional components are estimated conditionally on current values of parametric terms.
\end{enumerate}
To obtain inference for the regression parameters, we use a sandwich-type robust variance estimator that accounts for the correlation within subjects and the penalization on the spline coefficients. Specifically, the parameter vector $\pmb{\beta}$ corresponding to fixed effects (non-penalized) and the spline coefficient vectors $\pmb{\theta}$ and $\pmb{\theta}_c$ corresponding to the penalized complex carry-over effects are estimated by solving the penalized GEE described in equations \eqref{u1gee}–\eqref{u4gee}.

The penalization parameters $\lambda_c$, applied to the functional coefficients $\pmb{\theta}_c = (\theta_{c1}, \ldots, \theta_{cC})^\top$, control the inclusion and smoothness of the carry-over effects in the model. When all $\lambda_c$ values are very large, the corresponding penalization terms dominate the estimating equations, effectively shrinking all $\theta_{cj}$ coefficients toward zero. In this limiting case, the model discards the contribution of complex carry-over effects and reduces to a standard semiparametric GEE for repeated-measures crossover data, assuming no carry-over between periods.

Conversely, when all $\lambda_c$ values are set to zero, the model includes the full set of basis functions for carry-over effects, allowing the estimation of potentially complex carry-over patterns. However, this may lead to overfitting if the true carry-over structure is sparse or weak.

Therefore, the estimation of at least one $\theta_{c} \neq 0$ indicates the presence of at least one non-negligible carry-over effect in the data. Testing whether any functional coefficient associated with carry-over is significantly different from zero is a key step in assessing the need for modeling such effects. In practice, regularization allows balancing model complexity and parsimony, letting the data determine whether carry-over effects should be retained in the final model.
Because $\pmb{\beta}$ is not penalized, its variance estimator follows the classical sandwich form, while for the penalized spline parameters $\pmb{\theta}$ and $\pmb{\theta}_c$, the penalty matrices are incorporated. This ensures proper regularization is accounted for when constructing confidence bands and hypothesis testing.

The robust sandwich variance estimators for these parameters are given by:
\begin{align}
\widehat{\mathrm{Var}}(\hat{\pmb{\beta}}) &= 
\left( \sum_{i=1}^n \pmb{D}_{\beta i}^\top \pmb{V}_{3i}^{-1} \pmb{D}_{\beta i} \right)^{-1}
\left( \sum_{i=1}^n \pmb{D}_{\beta i}^\top \pmb{V}_{3i}^{-1} \pmb{r}_i \pmb{r}_i^\top \pmb{V}_{3i}^{-1} \pmb{D}_{\beta i} \right)
\left( \sum_{i=1}^n \pmb{D}_{\beta i}^\top \pmb{V}_{3i}^{-1} \pmb{D}_{\beta i} \right)^{-1} \label{sandubeta}
\\
\widehat{\mathrm{Var}}(\hat{\pmb{\theta}}) &= 
\left( \sum_{i=1}^n \pmb{D}_{\theta i}^\top \pmb{V}_{1i}^{-1} \pmb{D}_{\theta i} + \lambda_\theta \pmb{P}_\theta \right)^{-1}
\left( \sum_{i=1}^n \pmb{D}_{\theta i}^\top \pmb{V}_{1i}^{-1} \pmb{r}_i \pmb{r}_i^\top \pmb{V}_{1i}^{-1} \pmb{D}_{\theta i} \right)
\left( \sum_{i=1}^n \pmb{D}_{\theta i}^\top \pmb{V}_{1i}^{-1} \pmb{D}_{\theta i} + \lambda_\theta \pmb{P}_\theta \right)^{-1} 
\\
\widehat{\mathrm{Var}}(\hat{\pmb{\theta}}_c) &= 
\left( \sum_{i=1}^n \pmb{D}_{\theta_c i}^\top \pmb{V}_{2i}^{-1} \pmb{D}_{\theta_c i} + \lambda_c \pmb{P}_{\theta_c} \right)^{-1}
\left( \sum_{i=1}^n \pmb{D}_{\theta_c i}^\top \pmb{V}_{2i}^{-1} \pmb{r}_i \pmb{r}_i^\top \pmb{V}_{2i}^{-1} \pmb{D}_{\theta_c i} \right)
\left( \sum_{i=1}^n \pmb{D}_{\theta_c i}^\top \pmb{V}_{2i}^{-1} \pmb{D}_{\theta_c i} + \lambda_c \pmb{P}_{\theta_c} \right)^{-1}
\end{align}
where
$\pmb{D}_{\beta i} = \mathrm{diag}\left(\frac{\partial \mu_{ijk}}{\partial \pmb{\beta}}\right)_i, \quad
\pmb{D}_{\theta i} = \mathrm{diag}\left(\frac{\partial \mu_{ijk}}{\partial \pmb{\theta}}\right)_i,$ $\pmb{D}_{\theta_c i} = \mathrm{diag}\left(\frac{\partial \mu_{ijk}}{\partial \pmb{\theta}_c}\right)_i, \pmb{r}_i = \pmb{y}_i - \pmb{\mu}_i, \quad
\pmb{V}_{ji}$ is the variance-covariance matrix with working correlation structure for component $j$, $\lambda \pmb{P}, \quad \lambda_c \pmb{P}_{c}$ are the penalty matrices for $\pmb{\theta}$, and for each $\pmb{\theta}_c$,  respectively. 
\begin{theorem}\label{te2}
Let $\widehat{\pmb{\beta}}, \widehat{\pmb{\theta}}, \widehat{\pmb{\theta}}_c$ denote the estimators obtained from solving the penalized GEE system. Then,

\begin{enumerate}
    \item The estimating equations for non-penalized parameters are asymptotically unbiased:
    \[
    U_{\beta}(\widehat{\pmb{\beta}}) \xrightarrow{P} 0.
    \]

    \item The penalized score functions satisfy:
    \[
    \| U_{\theta}(\widehat{\pmb{\theta}}) + \lambda \pmb{P}_{1} \widehat{\pmb{\theta}} \| = o_P(1)
    \] 
    \[ 
    \| U_{\theta_c}(\widehat{\pmb{\theta}}_c) + \lambda_{c} \pmb{P}_{2} \widehat{\pmb{\theta}}_c \| = o_P(1), \quad c=1, \ldots, C 
    \]

    \item The estimator for the non-penalized parameter vector $\widehat{\pmb{\beta}}$ is asymptotically normal:
    \[
    \sqrt{n} (\widehat{\pmb{\beta}} - \pmb{\beta}_0) \xrightarrow{d} \mathcal{N}(0, \pmb{\Sigma}_{\beta}),
    \]
    where the asymptotic variance $\pmb{\Sigma}_{\beta}$ is given by the robust sandwich formula in Equation \eqref{sandubeta}.
\end{enumerate}
\end{theorem}
\begin{proof}
 See Appendix \ref{prooflema5}
\end{proof}
In summary, the proposed estimation procedure extends existing semiparametric GEE frameworks to accommodate general functional bases in a Hilbert space setting. By incorporating appropriate roughness penalties and data-driven selection of smoothing parameters, the methodology ensures stable and interpretable estimates of both time and complex carry-over effects in crossover designs with repeated measurements. 

In the following section, we assess the empirical performance of this approach through simulation studies designed to evaluate estimation accuracy, bias, and robustness under various scenarios.

\section{Simulation study}

To evaluate the finite-sample performance of the proposed estimator under different carry-over and time trend scenarios, we conducted a Monte Carlo simulation mimicking a typical two-period AB/BA crossover design with repeated measurements. For each subject, the outcome was generated as
\[
Y_{ijk} =\begin{bmatrix}
    1& x_{1ijk} & x_{2ijk}
\end{bmatrix}\begin{bmatrix}
    \beta_0 \\ \beta_1 \\\beta_2
\end{bmatrix} + f(Z_{ijk}) + f_{1}(Z_{ijk})+ f_{2}(Z_{ijk}) + \varepsilon_{ijk}\]

where $i=1,\ldots,n$ indexes experimental units, $j=1,2$ denotes the period, and $k=1,\ldots,L$ the measurement occasions within each period. The model includes the following components: $\beta_0$ is the global intercept, $\beta_1$ the effect of treatment B, and $\beta_2$ the effect of the second period. The indicator $x_{1ijk}$ equals 1 if experimental unit $i$ receives treatment B at time $k$, and 0 otherwise; similarly, $x_{2ijk}=1$ if $j=2$, and 0 otherwise. The variable $Z_{ijk}$ represents the measurement time within each period, with $Z_{ijk}=k$. The errors are assumed independent and normally distributed, $\varepsilon_{ijk} \sim N(0,\sigma^2)$. The smooth function $f(\cdot)$ captures the within-period time trend, $f_{1}(\cdot)$ represents the carry-over effect of treatment A on treatment B, and $f_{2}(\cdot)$ models a potential time-varying carry-over effect that may arise in the second period depending on the treatment sequence.

The true functional components were specified to reflect realistic nonlinear patterns. The time trend $f(t)$ follows a sinusoidal curve, while carry-over effects differ by sequence, being represented by quadratic or cosine functions. In particular, we consider
\begin{equation}
  f(Z_{ijk}) = \sin\!\Big(2\pi \frac{Z_{ijk}}{L}\Big),\quad f_{1}(Z_{ijk}) = \sin\!\Big(2\pi \frac{Z_{ijk}}{L}\Big) \text{ and, } f_{2}(Z_{ijk}) = \cos\!\Big(2\pi \frac{Z_{ijk}}{L}\Big),  
\end{equation}
where $L$ denotes the number of repeated measurements per period.
To ensure identifiability and fair comparison with models excluding carry-over effects, all functional components—$f(t)$, $f_{1}(t)$, and $f_{1}(t)$—were centered to have zero mean over the time domain. This centering avoids any systematic shift that could artificially favor the detection or estimation of carry-over effects, thus providing an unbiased and rigorous evaluation of the proposed methodology.
\begin{table}[!ht]
\centering
\begin{tabular}[t]{lllccc|ccc|ccc}
\toprule
\multicolumn{3}{c}{ } & \multicolumn{3}{c}{Estimate} & \multicolumn{3}{c}{Coverage} & \multicolumn{3}{c}{RMSE} \\
\cmidrule(l{3pt}r{3pt}){4-6} \cmidrule(l{3pt}r{3pt}){7-9} \cmidrule(l{3pt}r{3pt}){10-12}
$L$ & $n$ & Model & $\beta_1=-1$ & $\beta_1=0$ & $\beta_1=1$ & $\beta_1=-1$ & $\beta_1=0$ & $\beta_1=1$ & $\beta_1=-1$ & $\beta_1=0$ & $\beta_1=1$ \\
\midrule

% Bloque L=10, n=2
\multirow{6}{*}{10} & \multirow{6}{*}{2} & GAM-time & -1.001 & -0.001 & 1.010 & 0.974 & 0.966 & 0.966 & 1.044 & 1.044 & 1.041 \\
                    &                     & GEE-carry & -0.995 & 0.005 & 1.011 & 0.692 & 0.678 & 0.690 & 1.396 & 1.392 & 1.392 \\
                    &                     & GEE-int   & -0.995 & 0.005 & 1.011 & 0.692 & 0.678 & 0.690 & 1.213 & 1.209 & 1.207 \\
                    &                     & GEE-main  & -1.001 & -0.001 & 1.010 & 0.690 & 0.698 & 0.702 & 1.401 & 1.397 & 1.397 \\
                    &                     & GEE-smooth& -1.001 & -0.001 & 1.010 & 0.934 & 0.942 & 0.968 & 0.862 & 0.862 & 0.853 \\
                    &                     & GEE-spline& -1.001 & -0.001 & 1.010 & 0.834 & 0.822 & 0.818 & 0.862 & 0.862 & 0.853 \\
\cmidrule(lr){2-12}

% Bloque L=10, n=5
                    & \multirow{6}{*}{5} & GAM-time & -0.997 & 0.002 & 1.013 & 0.960 & 0.952 & 0.970 & 1.072 & 1.070 & 1.072 \\
                    &                     & GEE-carry & -1.001 & -0.001 & 1.013 & 0.900 & 0.884 & 0.878 & 1.407 & 1.405 & 1.407 \\
                    &                     & GEE-int   & -1.001 & -0.001 & 1.013 & 0.900 & 0.884 & 0.878 & 1.230 & 1.228 & 1.230 \\
                    &                     & GEE-main  & -0.997 & 0.002 & 1.013 & 0.884 & 0.846 & 0.880 & 1.409 & 1.407 & 1.408 \\
                    &                     & GEE-smooth& -0.997 & 0.002 & 1.013 & 0.958 & 0.956 & 0.966 & 0.948 & 0.946 & 0.947 \\
                    &                     & GEE-spline& -0.997 & 0.002 & 1.013 & 0.918 & 0.886 & 0.916 & 0.948 & 0.946 & 0.947 \\
\cmidrule(lr){2-12}

% Bloque L=10, n=10
                    & \multirow{6}{*}{10} & GAM-time & -1.002 & 0.005 & 1.003 & 0.966 & 0.962 & 0.966 & 1.080 & 1.083 & 1.078 \\
                    &                     & GEE-carry & -1.001 & 0.001 & 1.005 & 0.930 & 0.940 & 0.944 & 1.410 & 1.411 & 1.410 \\
                    &                     & GEE-int   & -1.001 & 0.001 & 1.005 & 0.930 & 0.940 & 0.944 & 1.236 & 1.236 & 1.234 \\
                    &                     & GEE-main  & -1.002 & 0.005 & 1.003 & 0.926 & 0.910 & 0.914 & 1.411 & 1.412 & 1.411 \\
                    &                     & GEE-smooth& -1.002 & 0.005 & 1.003 & 0.958 & 0.942 & 0.936 & 0.973 & 0.977 & 0.972 \\
                    &                     & GEE-spline& -1.002 & 0.005 & 1.003 & 0.938 & 0.932 & 0.936 & 0.973 & 0.977 & 0.972 \\
\midrule

% Bloque L=20, n=2
\multirow{6}{*}{20} & \multirow{6}{*}{2} & GAM-time & -0.990 & -0.001 & 0.988 & 0.966 & 0.972 & 0.954 & 1.068 & 1.066 & 1.066 \\
                    &                     & GEE-carry & -1.000 & 0.004 & 0.994 & 0.674 & 0.712 & 0.718 & 1.409 & 1.404 & 1.403 \\
                    &                     & GEE-int   & -1.000 & 0.004 & 0.994 & 0.674 & 0.712 & 0.718 & 1.209 & 1.212 & 1.207 \\
                    &                     & GEE-main  & -0.990 & -0.001 & 0.988 & 0.686 & 0.688 & 0.692 & 1.411 & 1.406 & 1.405 \\
                    &                     & GEE-smooth& -0.990 & -0.001 & 0.988 & 0.956 & 0.948 & 0.936 & 0.925 & 0.923 & 0.922 \\
                    &                     & GEE-spline& -0.990 & -0.001 & 0.988 & 0.816 & 0.818 & 0.806 & 0.925 & 0.923 & 0.922 \\
\cmidrule(lr){2-12}

% Bloque L=20, n=5
                    & \multirow{6}{*}{5} & GAM-time & -0.996 & 0.002 & 0.994 & 0.976 & 0.960 & 0.978 & 1.081 & 1.079 & 1.079 \\
                    &                     & GEE-carry & -1.003 & -0.002 & 0.996 & 0.890 & 0.878 & 0.860 & 1.411 & 1.409 & 1.408 \\
                    &                     & GEE-int   & -1.003 & -0.002 & 0.996 & 0.890 & 0.878 & 0.860 & 1.219 & 1.217 & 1.216 \\
                    &                     & GEE-main  & -0.996 & 0.002 & 0.994 & 0.878 & 0.870 & 0.888 & 1.412 & 1.409 & 1.409 \\
                    &                     & GEE-smooth& -0.996 & 0.002 & 0.994 & 0.956 & 0.958 & 0.944 & 0.971 & 0.968 & 0.968 \\
                    &                     & GEE-spline& -0.996 & 0.002 & 0.994 & 0.916 & 0.908 & 0.924 & 0.971 & 0.968 & 0.968 \\
\cmidrule(lr){2-12}

% Bloque L=20, n=10
                    & \multirow{6}{*}{10} & GAM-time & -1.002 & -0.005 & 1.003 & 0.976 & 0.968 & 0.960 & 1.085 & 1.082 & 1.085 \\
                    &                     & GEE-carry & -1.010 & -0.004 & 1.010 & 0.942 & 0.932 & 0.914 & 1.413 & 1.410 & 1.412 \\
                    &                     & GEE-int   & -1.010 & -0.004 & 1.010 & 0.942 & 0.932 & 0.914 & 1.222 & 1.219 & 1.221 \\
                    &                     & GEE-main  & -1.002 & -0.005 & 1.003 & 0.932 & 0.928 & 0.916 & 1.413 & 1.410 & 1.412 \\
                    &                     & GEE-smooth& -1.002 & -0.005 & 1.003 & 0.948 & 0.954 & 0.962 & 0.986 & 0.983 & 0.985 \\
                    &                     & GEE-spline& -1.002 & -0.005 & 1.003 & 0.938 & 0.944 & 0.932 & 0.986 & 0.983 & 0.985 \\
\bottomrule
\end{tabular}
\caption{Estimation performance for the treatment effect ($\beta_1 = \{-1, 0, 1\}$) across models. For each method, mean of estimates, empirical coverage, and the root mean squared error (RMSE) of the model, are reported across all simulations for $L=\{10,20,50\}$ and $n=\{2,5,10\}$.}
\label{tableSIM}
\end{table}

Scenarios varied the magnitude of the treatment effect $\beta_1$ from 0 to 10, while fixing the period effect at $\beta_2 = 0.2$. Each subject contributed observations across two periods, with the number of within-period time points set at $L \in \{10, 20, 50, 100\}$ to examine the impact of time series length on estimation. For each scenario, we simulated $n\in\{2, 5, 10,25,50\}$ subjects per sequence, assumed an exchangeable working correlation structure, and repeated the process 500 times.

%\caption{Estimation performance for the treatment effect ($\beta = -1$) across models. For each method, mean of estimates, empirical coverage, and the root mean squared error (RMSE) of the model, are reported across all simulations.}

Each simulated dataset was analyzed using six alternative models. The first model (GEE-main) is a basic generalized estimating equations (GEE) approach including only treatment and period main effects. The second model (GAM-time) extends this by incorporating a smooth time trend via a generalized additive model (GAM).
The third model (GEE-carry) adds a simple linear carry-over term to the baseline GEE. The fourth model (GEE-int) introduces an interaction between time and the linear carry-over effect, allowing for time-varying carry-over. The fifth model (GEE-smooth) is the proposed approach, which jointly estimates smooth effects for both time trends and carry-over functions using penalized splines. Finally, the sixth model (GEE-spline) fits the same smooth time and carry-over components as GEE-smooth but without any penalization, resulting in more flexible but potentially overfitted estimates.

Table~\ref{tableSIM} presents the simulation results for $n=\{2,5,10\}$ (number of experimental units per sequence) and $L=\{10,20,50\}$ (number of time points observed for each experimental unit within each period). All models exhibit accurate average estimation of the treatment effect $\beta_1$, regardless of whether its value is $-1$, $0$, or $1$. This is guaranteed by the fact that the AB/BA crossover design is balanced with respect to both periods and treatments, and also because, as previously mentioned, we used basis functions with mean zero to protect classical models from bias.

This behavior is also observed for more extreme values of $\beta_1$ tested in the simulation. When analyzing the coverage probabilities, we note that the GEE-based models without carry-over effects exhibit low coverage, with values around 70\%. This can be explained by the fact that, although the point estimates are unbiased, the standard errors of the estimators are highly sensitive to the variability induced by time and carry-over effects.

The GAM-time model shows good coverage levels, slightly above the nominal level of 95\%. However, since this model uses very strong penalization to avoid detecting complex carry-over effects, the variance estimation of the treatment effect tends to be slightly inflated. The model with simple carry-over effects has a coverage slightly below 0.95 but still clearly outperforms the standard GEE models. Our proposed model with complex carry-over effects (unpenalized) achieves coverage levels close to 95\% when either $n$ or $L$ increases. The inclusion of penalization improves the performance for large values of $L$, correcting slight undercoverage.

Regarding the RMSE column for the residual error, we observe very low values across the proposed models, and close to $1$, which was the value of the true error variance $\sigma^2=1$ used in all simulations. Therefore, the proposed methodology provides unbiased estimation of the residual variability.

\begin{figure}[ht]
\centering 
\includegraphics[width=15cm]{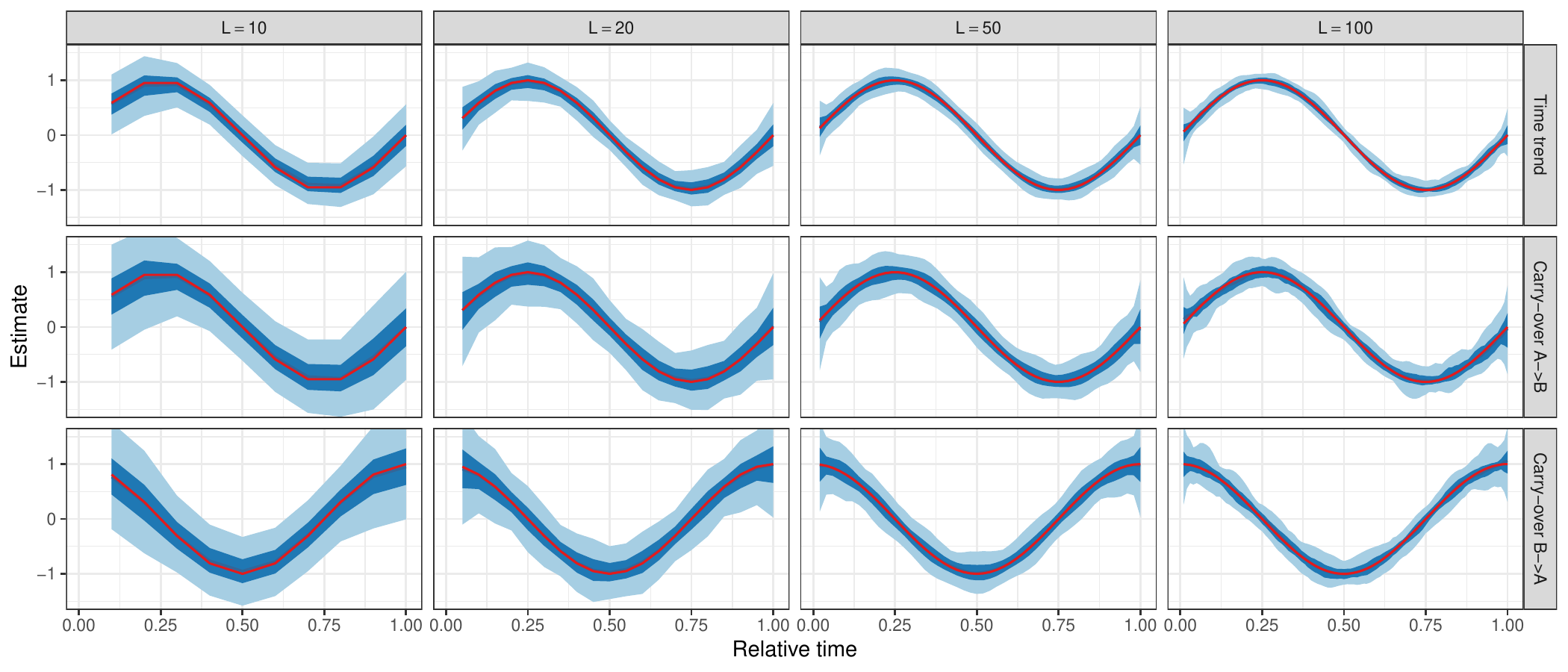}
\caption{Confidence bands and median of estimated functions by number of time points ($L$) and effect type with $n=5$ experimental units by sequence. Light and medium blue areas show the 95\% and 50\% quantile intervals, respectively. The red line is the true function.}
\label{first}
\end{figure}

\begin{figure}[ht]
\centering 
\includegraphics[width=15cm]{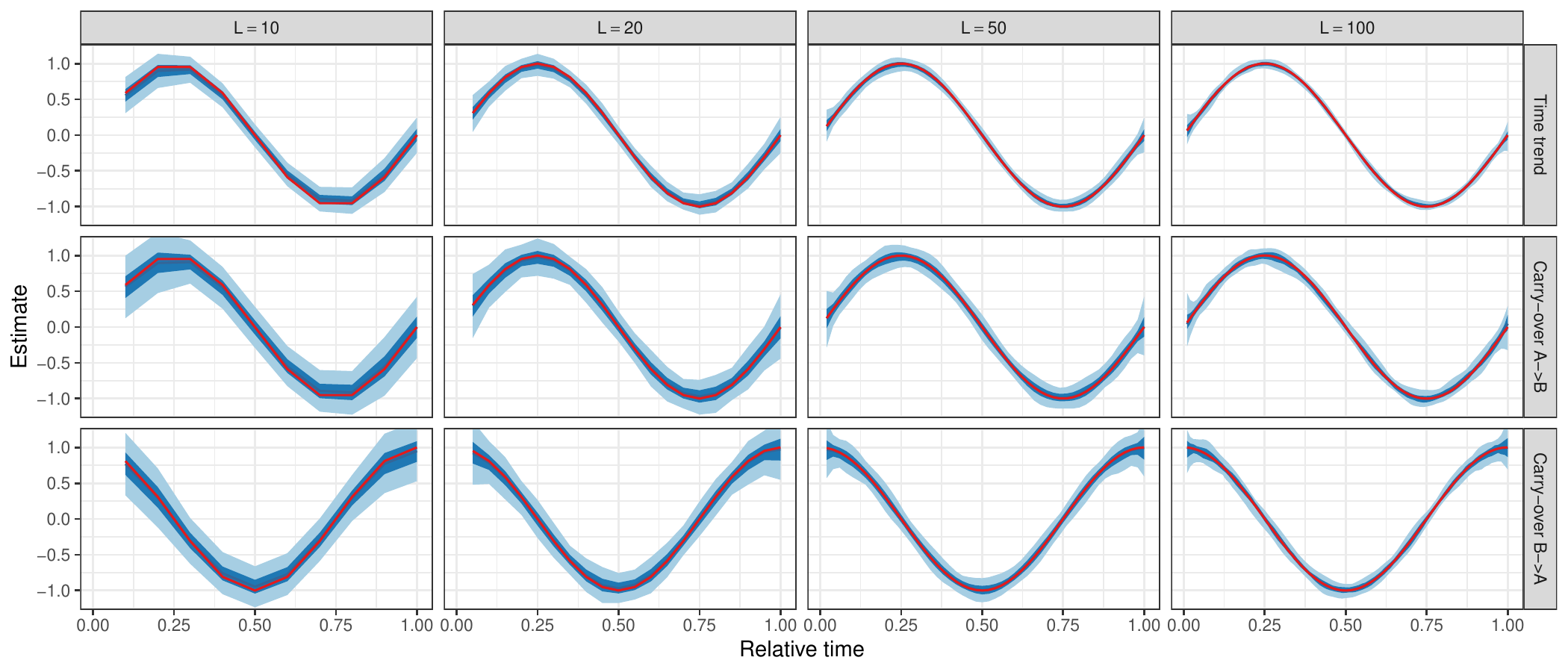}
\caption{Confidence bands and median of estimated functions by number of time points ($L$) and effect type with $n=25$ experimental units by sequence. Light and medium blue areas show the 95\% and 50\% quantile intervals, respectively. The red line is the true function.}
\label{second}
\end{figure}
Figures~\ref{first} and \ref{second} show the estimation of the functional effects for time and for the two complex carry-over effects in the proposed model, for $n=5$ and $n=10$, respectively. It is evident that the confidence bands for the time effect are much narrower, since this effect is estimated using twice as much information as the carry-over effects. In particular, the time effect is informed by data from both periods, while the carry-over effects are only informed by data from the second period.

Additionally, it can be seen that as the number of time points increases, the estimated curves concentrate more closely around the true functions, indicating improved estimation accuracy. The shading around the estimated curves reflects the variability across simulation runs. These results support the consistency established in Theorem 2, and illustrate how the penalization approach enables smooth estimation even when using 100 time points.

These patterns are consistent across all other simulation scenarios considered, including different values of $n$, $L$, and treatment effects. Due to space limitations, we do not present all results here. However, the omitted results exhibit similar behavior, reinforcing the robustness and generalizability of the proposed methodology under a wide range of realistic experimental settings. The full set of simulation tables and graphical summaries is provided in the supplementary material.

\section{Application}

\begin{figure}[ht]
\centering 
\includegraphics[width=14cm]{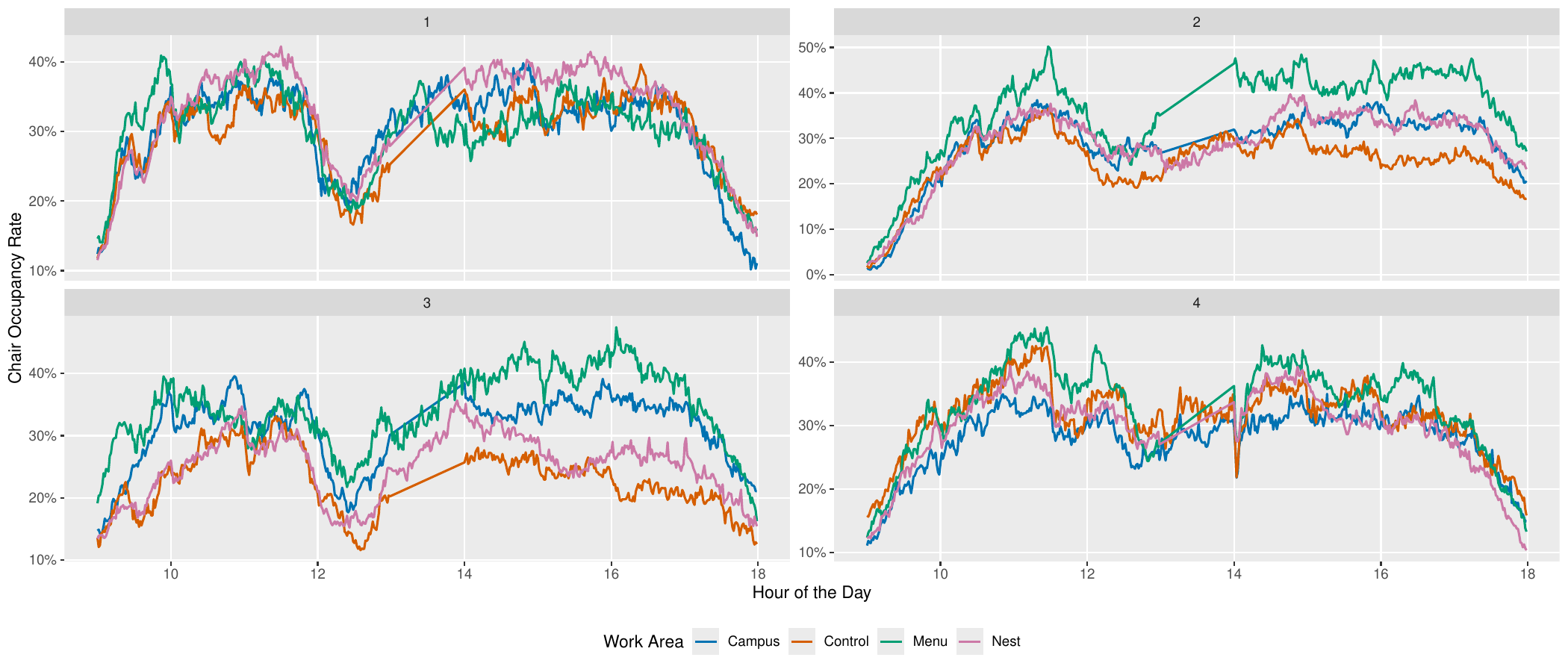}
\caption{Average chair occupancy rate over the course of the day, stratified by group, work area (treatment), and wave (period). Each line represents the average proportion of time that chairs were detected as occupied at each minute of the day. }
\label{datosreales1}
\end{figure}

To evaluate the performance of the proposed semiparametric penalized estimation methodology in a real crossover setting, we analyze data from a workplace experiment conducted by Booking.com and described in \citet{pitchforth2020work}. The study implemented a Williams design with 288 ($n=72$) participants assigned to four groups, each exposed to a different sequence of four office layouts: A (Menu), B (Control), C (Nest), and D (Campus). Each treatment period lasted two weeks. Seat occupancy was originally recorded at a high frequency of four observations per minute using infrared sensors. For this application, the data were aggregated and summarized at 5-minute intervals, resulting in $L=96$ observations per period per participant. This aggregation provided a detailed yet manageable temporal resolution to model occupancy dynamics over the two-week treatment periods. 

The structure of the crossover design is presented in Table~\ref{tableZA}. Each group followed a unique sequence of treatments across the four periods, ensuring balance and control over period and sequence effects.

\begin{table}[ht]
    \centering
   \begin{tabular}{c|c|c|c|c|}
        Sequence & Period 1 & Period 2 & Period 3 & Period 4\\
        \hline
         Group 1 & B & A & D & C \\
         Group 2 & C & D & A & B \\
         Group 3 & D & B & C & A \\
         Group 4 & A & C & B & D \\
        \hline
    \end{tabular}
    \caption{Structure of the crossover design in work environment experiment}
    \label{tableZA}
\end{table}

Figure~\ref{datosreales1} displays the empirical average proportion of time that each chair was occupied throughout the day, stratified by treatment. These descriptive curves suggest strong temporal patterns and potential carry-over effects from previous treatments, motivating a flexible modeling approach.
Several models were fitted to the dataset under a logistic link and binomial distribution. The proposed methodology allows for the estimation of simple and complex carry-over effects using penalized functional components. In this analysis, we fitted and compared several models incorporating different types of carry-over effects: (i) complex carry-over effects estimated with penalized splines (penalized complex), (ii) simple carry-over effects estimated with penalized splines (penalized simple), (iii) linear carry-over effects (simple linear time and carry-over), and (iv) a model without any carry-over effects (without). Working correlation structures considered were independence and autoregressive.

\begin{table}[ht]
\centering
\begin{tabular}{|c|c|c|}
\hline
Model & Correlation & QIC \\
\hline
Complex Penalized & Autoregressive & 2960.30 \\
Complex Penalized & Independence & 2908.11\\
Simple Penalized & Autoregressive & 2949.12 \\
Simple Penalized & Independence & 2938.87\\
Linear Carry-over & Autoregressive & 58831.94\\
Linear Carry-over & Independence & 58773.25\\
Without Carry-over & Autoregressive & 58692.16\\
Without Carry-over & Independence & 58664.20\\
\hline
\end{tabular}
\caption{QIC for the fitted models to the proportions of times the chair is occupied}
\label{tabla800}
\end{table}
\begin{figure}[ht]
\centering 
\includegraphics[width=16cm]{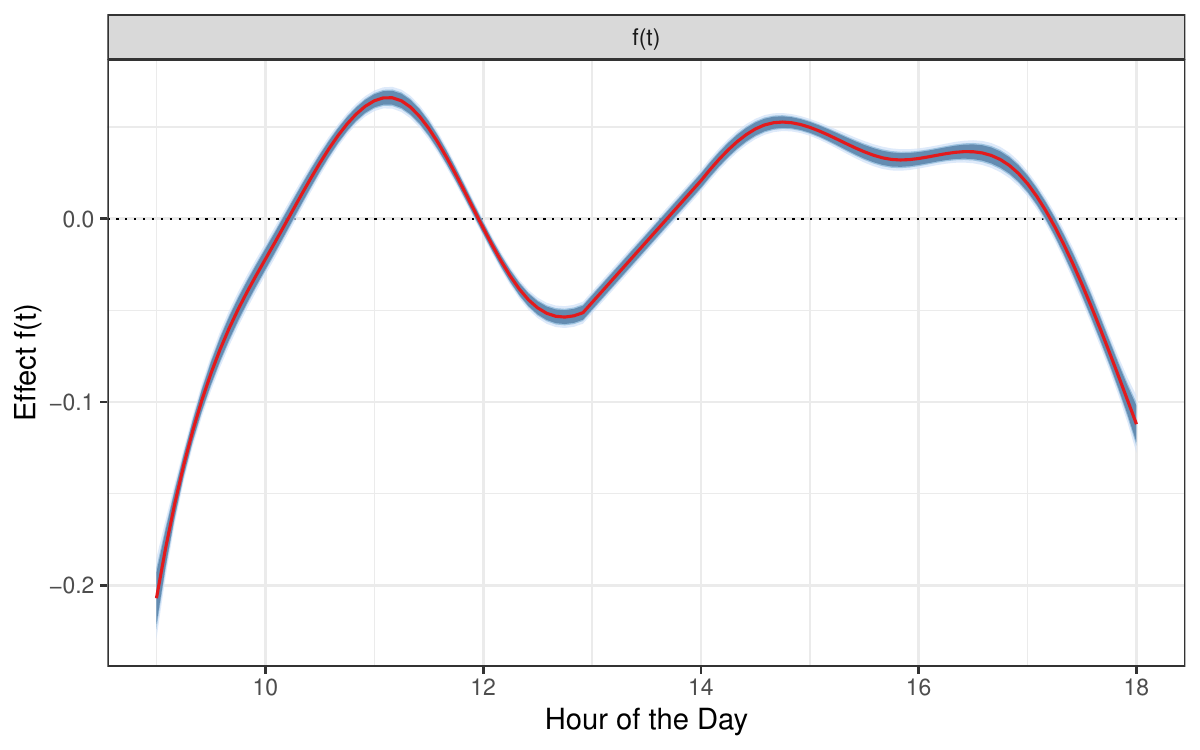}
\caption{Estimated smooth time effects. Estimation is based on penalized generalized estimating equations with functional terms.}
\label{TEf}
\end{figure}
As shown in Table~\ref{tabla800}, the semiparametric penalized GEE model incorporating complex carry-over effects and an independence working correlation structure achieved the lowest $QIC$, indicating superior fit and parsimony. This result highlights the ability of the proposed methodology to flexibly estimate carry-over effects, capturing nuanced temporal dependencies that simpler or parametric carry-over specifications fail to model effectively.
A grid of penalization parameters ($\lambda$) was tested, ranging from very small to very large values $\lambda \in \{0, 0.01, 0.1, 1, 10, 10^2, 10^3, \ldots, 10^9\}$.
The optimal $\lambda$ values for each smooth term were selected by minimizing the Quasi-likelihood under the Independence model Criterion (QIC). Table~\ref{tab:lambda_selection} summarizes the selected penalty values for the temporal and carry-over smooth effects.
\begin{table}[ht]
\centering
\small
\begin{tabular}{|rl|rl|}
\toprule
$\lambda$ & Carry-over effect & $\lambda$ & Carry-over effect \\
\midrule
$10^{8}$ & Campus $\to$ Control & 10         & Control $\to$ Menu \\
$10^{4}$ & Campus $\to$ Menu    & $10^{4}$ & Control $\to$ Nest \\
$10^{5}$ & Campus $\to$ Nest    & $10^{2}$ & Menu $\to$ Campus \\
$10^{6}$ & Control $\to$ Campus & $10^{3}$ & Menu $\to$ Control \\
$10^{6}$ & Menu $\to$ Nest     & $10^{6}$ & Nest $\to$ Campus \\
$10^{6}$ & Nest $\to$ Control  & $10^{5}$ & Nest $\to$ Menu \\
\bottomrule
\end{tabular}
\caption{Penalization parameters $\lambda$ selected by QIC minimization for the carry-over effects. The smoothing parameter for the time effect was selected at $\lambda = 0.01$.}
\label{tab:lambda_selection}
\end{table}
The selected penalization parameters reflect the degree of smoothness imposed on each functional effect. Lower $\lambda$ values correspond to less penalization, allowing more flexibility and variability in the estimated effect, as observed for the temporal effect ($\lambda = 0.01$). Conversely, very large $\lambda$ values (up to $10^8$) indicate strong smoothing, effectively shrinking the corresponding carry-over effects towards simpler functions or even zero.

This pattern suggests that the main temporal trend in seat occupancy requires flexible modeling to capture its dynamics accurately, while most carry-over effects exhibit limited complexity and are heavily regularized. The variation in $\lambda$ across different carry-over transitions also points to heterogeneous delayed influences between treatments, with some carry-overs necessitating more smoothing to avoid overfitting.

The final selected model is given by:
\begin{align*}
\ln\Bigg(\frac{p_{ijk}}{1 - p_{ijk}}\Bigg) &= \pmb{x}_{ijk}^\top \pmb{\beta} + f(Z_{ijk}) + \sum_{c=1}^{12} f_c(Z_{ijk}),
\end{align*}
where $\pmb{x}_{ijk}$ contains the period and treatment variables, $f_1$ captures the overall temporal trend, and $f_1, \dots, f_{12}$ represent the complex first-order carry-over effects, all estimated with penalized splines.

\begin{figure}[ht]
\centering 
\includegraphics[width=16cm]{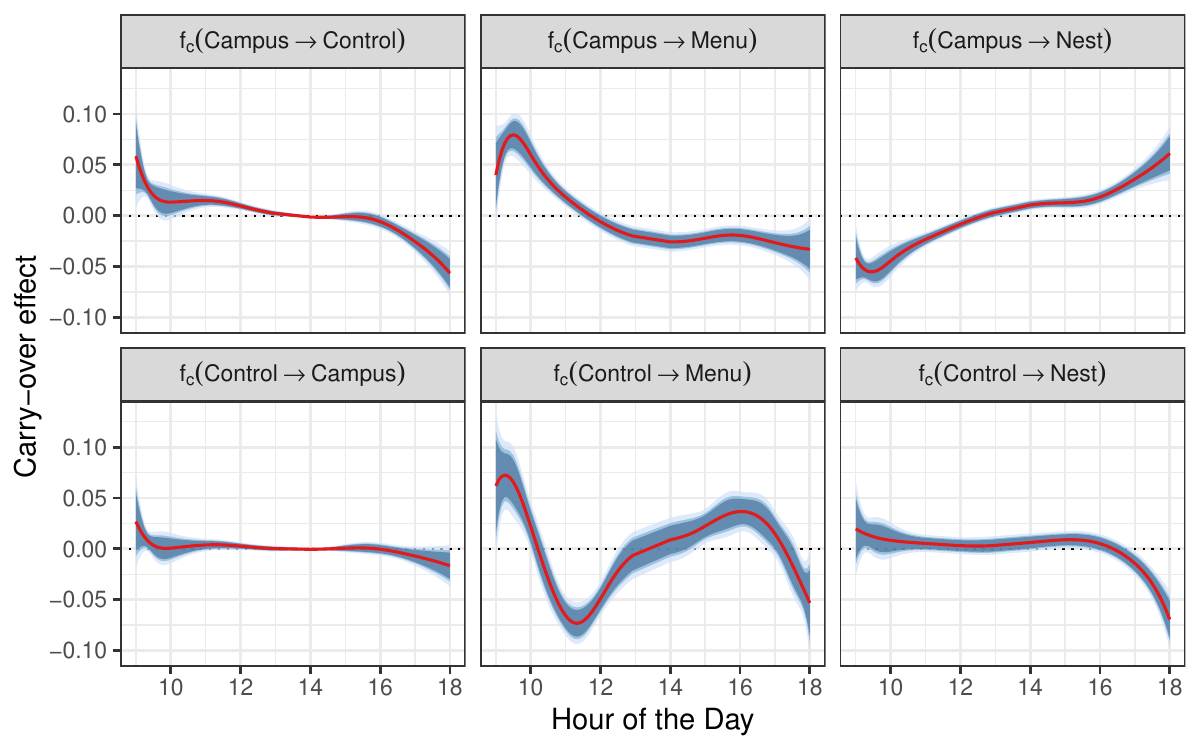}
\caption{Estimated smooth carry-over effects when transitioning from the \textit{Campus} treatment to each of the remaining treatments (\textit{Control}, \textit{Menu}, and \textit{Nest}) in the first row, and analogously from the \textit{Control} treatment in the second row. Each subplot corresponds to a specific treatment sequence (e.g., \textit{Campus}~$\rightarrow$~\textit{Control}). Estimation is based on penalized generalized estimating equations with functional terms.}
\label{CarryA}
\end{figure}

\begin{figure}[ht]
\centering 
\includegraphics[width=16cm]{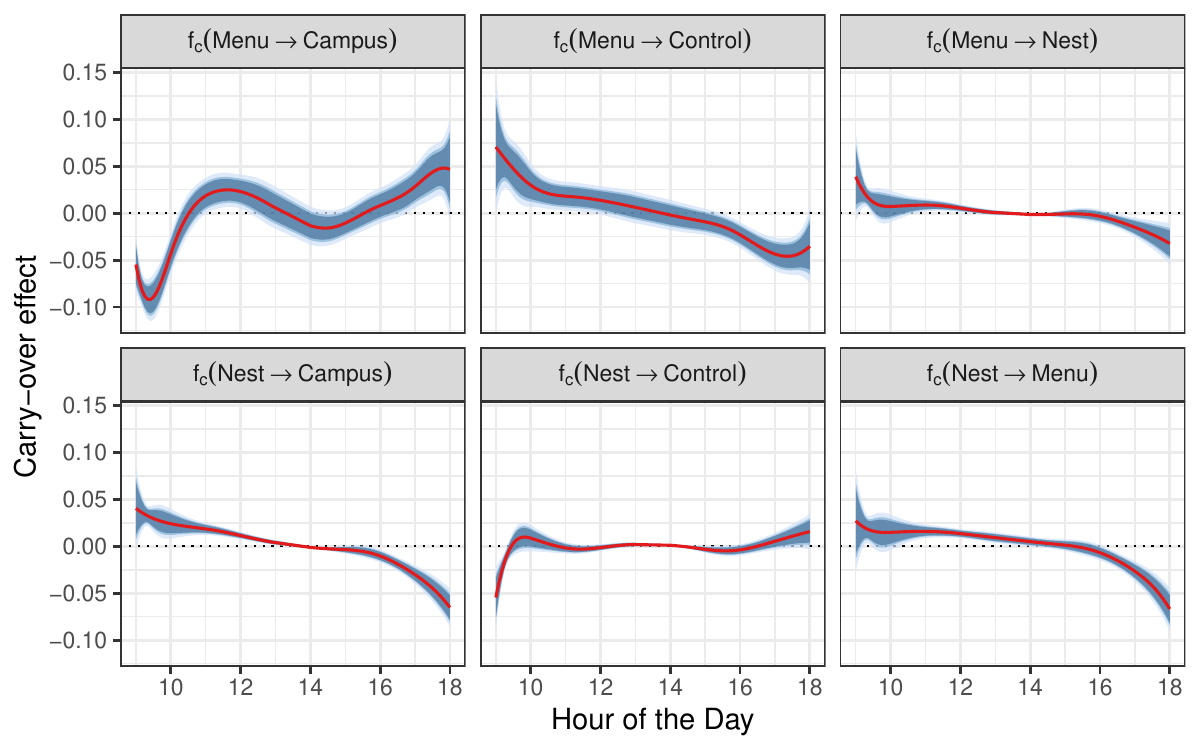}
\caption{Estimated smooth carry-over effects when transitioning from the \textit{Menu} treatment to each of the remaining treatments (\textit{Control}, \textit{Campus}, and \textit{Nest}) in the first row, and analogously from the \textit{Nest} treatment in the second row.}
\label{CarryB}
\end{figure}

The three figures \ref{TEf}, \ref{CarryA}, and \ref{CarryB} display confidence bands at the 90\%, 95\%, and 99\% levels. The time effect closely follows the pattern observed in the real data. The confidence bands are very narrow due to the large sample size (\(n=288\)) and the 96 observations used to estimate the curves. As seen in the simulation study, this results in very small error bands. A local minimum is observed around lunchtime, along with a bimodal pattern featuring peaks around 11 AM and 3 PM.

Figure~\ref{CarryA} shows the carry-over effects from the Campus and Control treatments. The effect of Campus on Control is decreasing but close to zero, similar to its effect on Menu, while it differs markedly from its effect on Nest. This supports the interpretation that the carry-over effect of Campus is complex rather than simple. The interpretability gain arises because participants receiving Nest after Menu exhibit an increasing occupancy effect as the day progresses, in contrast to those who received Menu after Control, where occupancy decreases over the day.

The figure also depicts the effect of Control on the other treatments. It is nearly zero for Campus and Nest but significant and highly variable on Menu, where it decreases early in the day and approaches zero after lunch. Similarly, in Figure~\ref{CarryB}, the effect of Menu on Nest is close to zero, but significant and decreasing on Control and increasing on Campus. The effect of Nest on Control is almost zero, except for a decreasing effect on Campus, similar to its effect on Menu, although that is very close to zero.

In summary, the estimated smooth carry-over effects reveal complex temporal dependencies between treatments that simple carry-over models cannot capture. The rich longitudinal data and the penalized semiparametric modeling approach allow for a nuanced understanding of how prior office designs influence subsequent occupancy patterns throughout the day. This highlights the importance of accounting for complex carry-over effects in crossover designs analyzing behavioral time series data.

\section{Conclusions}
The discussion on complex carry-over effects has been long-standing in the field of statistics applied to clinical and human research. Until now, no methodology has allowed for the consistent estimation of such effects, which has likely contributed to their frequent omission in many studies, either by assuming they do not exist or by modeling only oversimplified carry-over structures. The former assumption received some support from pharmacological studies, while the latter was primarily defended by statisticians seeking mathematical tractability rather than being grounded in empirical evidence.

This work contributes in two fundamental ways. First, it establishes the conditions under which complex carry-over effects are estimable in crossover designs. Second, it introduces a penalized GEE framework capable of consistently estimating these effects when they are present. The penalization mechanism plays a dual role: with large penalty parameters, carry-over effects are shrunk toward zero when absent; with moderate penalties, their true magnitude is recovered when relevant. This adaptive property ensures both robustness and flexibility, reducing bias in treatment effect estimation while guarding against overfitting.

Simulation studies confirm that the proposed methodology achieves a balance between accuracy and parsimony across diverse scenarios, and the real-data application demonstrates its practical relevance. Rather than treating carry-over effects as nuisance parameters to be ignored, the framework provides a principled way to incorporate them into analysis, offering richer scientific insight and supporting more effective treatment design.

Overall, by first defining the conditions of estimability and then proposing a semiparametric penalized spline approach, this work closes a long-standing methodological gap in crossover trial analysis. It enhances the validity and interpretability of crossover studies and opens new avenues for nuanced and reliable modeling of treatment effects in applied research.

\section*{Supplementary files}
\textbf{Supplementary file 1}: {An R code with the simulation exercise and the data and their analysis of work environment levels }

\appendix
\section{Proofs}
\subsection{Proof of Lemma \ref{lema1}}\label{prooflema1}
\begin{proof}
Since $\mathbf{1}_{n} \notin \mathcal{C}(\pmb{A})$ then $\nexists \, \pmb{x}\in \mathbb{R}^{p}$ such that
    \begin{align*}
        \pmb{\mathcal{A}}\pmb{x}&= \mathbf{1}_{n} \\
        \mathbf{1}_L\otimes \pmb{\mathcal{A}}\pmb{x}&= \mathbf{1}_L\otimes\mathbf{1}_{n}\\
        (\mathbf{1}_L\otimes \pmb{\mathcal{A}})\pmb{x}&= \mathbf{1}_{nL}
    \end{align*}
then $\mathbf{1}_{nL} \notin \mathcal{C}(\mathbf{1}_L\otimes \pmb{\mathcal{A}})$. Furthermore, $\mathbf{1}_L\otimes \pmb{A}$ is a permutation of rows of the matrix $\pmb{A}\otimes \mathbf{1}_L$.  Therefore, $(\mathbf{1}_L\otimes \pmb{\mathcal{A}})\pmb{x}= \mathbf{1}_{nL}$ is consistent if and only if $(\pmb{A}\otimes \mathbf{1}_L)\pmb{y}= \mathbf{1}_{nL}$ is consistent, with $\pmb{x}, \pmb{y} \in \mathbb{R}^{p}$, so $\mathbf{1}_{nL} \notin \mathcal{C}(\mathbf{1}_L\otimes \pmb{A})$.
\end{proof}
\subsection{Proof of Lemma \ref{lema2}}\label{prooflema2}
\begin{proof}
 Let $\mathbf{a}_j$ denote the $j$-th column of the matrix $\mathbf{A}$, and similarly let $\mathbf{b}_j$ be the $j$-th column of $\mathbf{B}$. Since
\[
\operatorname{rank}([\mathbf{A} \,\vert\, \mathbf{B}]) = p + q,
\]
it follows that $\mathbf{a}_j \notin \mathcal{C}(\mathbf{B})$ and $\mathbf{b}_j \notin \mathcal{C}(\mathbf{A})$ for all $j$. Now, consider the Kronecker products $\mathbf{a}_j \otimes \mathbf{1}_L$ and $\mathbf{b}_j \otimes \mathbf{1}_L$, where each column of the matrix $\mathbf{A} \otimes \mathbf{1}_L$ is of the form
\[
\mathbf{a}_j \otimes \mathbf{1}_L = 
\begin{bmatrix}
a_{1j} \mathbf{1}_L \\
a_{2j} \mathbf{1}_L \\
\vdots \\
a_{nj} \mathbf{1}_L
\end{bmatrix}
\in \mathbb{R}^{nL}.
\]
An analogous expression holds for the columns of $\mathbf{B} \otimes \mathbf{1}_L$. By standard properties of the Kronecker product (e.g., Horn and Johnson, 1991), linear independence is preserved. That is, if $\mathbf{a}_j \notin \mathcal{C}(\mathbf{B})$, then $\mathbf{a}_j \otimes \mathbf{1}_L \notin \mathcal{C}(\mathbf{B} \otimes \mathbf{1}_L)$, and similarly for $\mathbf{b}_j$.

Therefore, the columns of $\mathbf{A} \otimes \mathbf{1}_L$ remain linearly independent from those of $\mathbf{B} \otimes \mathbf{1}_L$, and then
\[
\operatorname{rank}([\mathbf{A} \otimes \mathbf{1}_L \,\vert\, \mathbf{B} \otimes \mathbf{1}_L]) = p + q.
\]
\end{proof}
\subsection{Proof of Lemma \ref{lema3}}\label{prooflema3}

\begin{proof}
The form of the matrix $\pmb{D}\otimes\mathbf{1}_{L}$ is: 
    \begin{equation*}
        \pmb{D}\otimes\mathbf{1}_{L} = \begin{bmatrix}
            d_{11}\mathbf{1}_{L} & d_{12}\mathbf{1}_{L} & \cdots & d_{1r}\mathbf{1}_{L}\\
            d_{21}\mathbf{1}_{L} & d_{22}\mathbf{1}_{L} & \cdots & d_{2r}\mathbf{1}_{L}\\
            \vdots & \vdots & \ddots & \vdots\\
            d_{n1}\mathbf{1}_{L} & d_{n2}\mathbf{1}_{L} & \cdots & d_{nr}\mathbf{1}_{L}
        \end{bmatrix}
    \end{equation*}
    and the matrix $\pmb{B} \otimes \pmb{A}$ has the form: 
    \begin{equation*}
        \pmb{B}\otimes\pmb{A} = \begin{bmatrix}
            b_{11}\pmb{A} & b_{12}\pmb{A} & \cdots & b_{1p}\pmb{A}\\
            b_{21}\pmb{A} & b_{22}\pmb{A} & \cdots & b_{2p}\pmb{A}\\
            \vdots & \vdots & \ddots & \vdots\\
            b_{n1}\pmb{A} & b_{n2}\pmb{A} & \cdots & b_{np}\pmb{A}
        \end{bmatrix}
    \end{equation*}
  and since $d_{ij}$ and $b_{ij}$ can only take the value of $0$ or $1$. If $d_{ij}=1$, then, $d_{ij}\mathbf{1}_{L}\notin \mathcal{C}(b_{ij}\pmb{A})$ and also,
$d_{i'j'}\mathbf{1}_{L} \notin \mathcal{C}([b_{i1}\pmb{A},  \cdots, b_{ij}\pmb{A} \, \cdots\ b_{ip}\pmb{A}])$. Furthermore, since $rank(\pmb{D})=r$, then $\sum_{i=1}^nd_{ij}\geq 1$, therefore, for any $j\in \{1, \ldots , r\}$
$[d_{1j}\mathbf{1}^\top_{L}, \ldots, d_{nj}\mathbf{1}^\top_{L}]^\top \notin \mathcal{C}(\pmb{B}\otimes\pmb {A})$. That is, the column space of $\pmb{B}\otimes\pmb{A}$ is linearly independent of the column space of $\pmb{D}\otimes\mathbf{1}_{L }$.
Then, $rank([\pmb{D}\otimes\mathbf{1}_{L}\,\vdots \pmb{B} \otimes \pmb{A}])=rank(\pmb{D}\otimes\mathbf{1}_{L})+ rank(\pmb{B} \otimes \pmb{A})=r\times 1 +p\times q=r+pq$.
\end{proof}

\subsection{Proof of Theorem \ref{te1}}\label{prooflema4}
\begin{proof}
Consider the full design matrix
\[
\pmb{X} = \left[\mathbf{1}_{nPL} \quad \pmb{T} \otimes \mathbf{1}_L \quad \pmb{\mathcal{P}} \otimes \mathbf{1}_L \quad \mathbf{1}_{nP}\otimes \pmb{\Phi}  \quad \pmb{C} \otimes\pmb{\Phi}\right]_{nPL \times q},
\]
where $\mathbf{1}_{nPL}$ is the intercept, $\pmb{T}$ and $\pmb{\mathcal{P}}$ are treatment and period design matrices of dimensions $nP \times (D-1)$ and $nP \times (P-1)$ respectively,  $\mathbf{1}_L$ is an $L \times 1$ vector of ones, $\pmb{\mathcal{C}}$ is the carry-over indicator matrix,  $\pmb{\Phi} \in \mathbb{R}^{L \times d}$ is the functional basis evaluation matrix at the repeated times $t_k$, and $q=1 + (D-1) + (P-1) + (C+1)\cdot d$.

Under standard identifiability restrictions for treatment and period (e.g., $\sum_d \tau_d = 0$, $\sum_j \pi_j = 0$), and under Assumptions \ref{as1} and \ref{as2}, $\pmb{T}$ and $\pmb{\mathcal{P}}$ are known to be of full column rank and linearly independent (\cite{senn1992simple}, and \cite{ken15}).

By Lemma \ref{lema2}, it is clear that:
\[
\mathrm{rank}([\pmb{T} \otimes \mathbf{1}_L \quad \pmb{\mathcal{P}} \otimes \mathbf{1}_L]) = \mathrm{rank}(\pmb{T}) + \mathrm{rank}(\pmb{\mathcal{P}}).
\]
Let $\pmb{C}$ be an indicator matrix of dimension $nP \times C$ with carry-over effects, under Assumption \ref{as3} has full column rank (one indicator per carry-over effect, see \citet{cruz2023correlation}). Under Assumption \ref{as5}, $\pmb{\Phi}$ has full column rank $d$ if the measurement times are distinct and suitable. 
By properties of Kronecker products, the carry-over design matrix is $\pmb{C} \otimes \pmb{\Phi}$ satisfying:
\[
\mathrm{rank}(\pmb{C} \otimes \Phi) = \mathrm{rank}(\pmb{\mathcal{C}}) \cdot \mathrm{rank}(\pmb{\Phi}) = C \cdot d.
\]

Since $\mathbf{1}_{nPL}$, $\pmb{T} \otimes \mathbf{1}_L$, $\pmb{\mathcal{P}} \otimes \mathbf{1}_L$, and $\pmb{C} \otimes \Phi$ correspond to intercept, treatment, period, and carry-over effects, respectively, and by Lemmas \ref{lema1}, \ref{lema2}, and \ref{lema3}, their column spaces are mutually linearly independent under the design assumptions. So, the overall design matrix $\pmb{X}$ has full column rank equal to:
\[
1 + (D-1) + (P-1) + (C+1) \cdot d.
\]
By the classical theory of linear models \cite{mcculloch2004generalized}, the parameter vector of coefficients (including $\pmb{\theta}_c$) is estimable if and only if the design matrix $\pmb{X}$ has full column rank.

Thus, the two stated conditions are necessary and sufficient for estimability of the complex carry-over effects modeled via the basis $\Phi$.

\end{proof}
\subsection{Proof of Theorem \ref{te2}}\label{prooflema5}
For the construction of the design matrix $\mathbf{X}$ in Equation \eqref{xxx}, the following conditions hold:

\begin{enumerate}
    \item The covariates $\mathbf{x}_{ijk}$, with $1 \leq i \leq n$, $1 \leq j \leq P$, and $1 \leq k \leq L$, are uniformly bounded.  

    \item From Theorem \ref{te1}, it follows that $\pmb{\beta}$ belongs to a compact subset $\mathcal{B}\subset \mathbb{R}^{\dim(\pmb{\beta})}$. Moreover, if the model in Equation \eqref{ectruth} is correctly specified, then the estimator $\hat{\pmb{\beta}}$ in Equation \eqref{u3gee_pen} lies in the interior of $\mathcal{B}$, by Theorem 1 of \citet{liang1986longitudinal}.  

    \item From Theorem \ref{te1}, there exist finite positive constants $b_1$ and $b_2$ such that  
    \[
    b_1 \leq \lambda_{\min}\left(\frac{1}{n}\sum_{i=1}^n \mathbf{X}_i^\top \mathbf{X}_i\right) \leq 
    \lambda_{\max}\left(\frac{1}{n}\sum_{i=1}^n \mathbf{X}_i^\top \mathbf{X}_i\right) \leq b_2,
    \]  
    where $\lambda_{\min}$ and $\lambda_{\max}$ denote, respectively, the minimum and maximum eigenvalues of the matrix $\frac{1}{n}\sum_{i=1}^n \mathbf{X}_i^\top \mathbf{X}_i$.  

    \item The estimated correlation matrix $\pmb{R}(\hat{\pmb{\alpha}})$ obtained from Equation \eqref{u4gee} has eigenvalues bounded away from both zero and $+\infty$ \citep{Davis}. Moreover, by Theorem 1 of \citet{balan2005asymptotic}, it satisfies  
    \[
    \| \pmb{R}(\hat{\pmb{\alpha}})^{-1} - \pmb{R}(\pmb{\alpha})^{-1} \| = \mathcal{O}_p\left(\sqrt{\tfrac{s_n}{n}}\right),
    \]  
    where $s_n$ is defined in Equation (5) of \citet{wang2012penalized}.  

    \item From Lemma 2 of \citet{xie2003asymptotics}, there exists a constant $M_1 > 0$ and $\delta > 0$ such that  
    \[
    \mathbb{E}\!\left(\|\epsilon_{i}(\pmb{\beta})\|^{2+\delta}\right)\leq M_1,
    \]  
    where $\epsilon_{i}(\pmb{\beta})=\pmb{V}_{3i}^{-1/2} \left(\pmb{y}_i - \pmb{\mu}_i[\pmb{\beta}, \pmb{\theta}, \pmb{\theta}_c] \right)$ as defined in Equation \eqref{u3gee_pen}.  
    Additionally, since $Y_{ijk}$ belongs to the exponential family, there exist positive constants $M_2$ and $M_3$ such that  
    \[
    \mathbb{E}\!\left(\exp \left(M_2 \, |\epsilon_{ijk}(\pmb{\beta})| \right)\right)\leq M_3,
    \]  
    where $\epsilon_{ijk}(\pmb{\beta})$ is the $jk$-th component of $\epsilon_{i}(\pmb{\beta})$.
\end{enumerate}
From 1) to 5), it is possible to apply Theorem 1) of \citet{wang2012penalized} over the $\hat{\pmb{\beta}}$ obtained from Equations \eqref{penU1} and \eqref{penU2}, and it is obtained that  
  \[
    \| U_{\theta}(\widehat{\pmb{\theta}}) + \lambda \pmb{P}_{1} \widehat{\pmb{\theta}} \| = o_P(1)
    \] 
    \[ 
    \| U_{\theta_c}(\widehat{\pmb{\theta}}_c) + \lambda_{c} \pmb{P}_{2} \widehat{\pmb{\theta}}_c \| = o_P(1), \quad c=1, \ldots, C 
    \]

And using 1) to 5), Theorem 1) of \citet{CruzSemi2023} and Theorem 3) of \citet{lu2024penalized} it is obtained that 
\[
    \sqrt{n} (\widehat{\pmb{\beta}} - \pmb{\beta}_0) \xrightarrow{d} \mathcal{N}(0, \pmb{\Sigma}_{\beta}),
    \]


\begin{thebibliography}{31}
\expandafter\ifx\csname natexlab\endcsname\relax\def\natexlab#1{#1}\fi
\expandafter\ifx\csname url\endcsname\relax
  \def\url#1{\texttt{#1}}\fi
\expandafter\ifx\csname urlprefix\endcsname\relax\def\urlprefix{URL }\fi

\bibitem[{Balan and Schiopu-Kratina(2005)}]{balan2005asymptotic}
Balan, R.M. \& Schiopu-Kratina, I. (2005) Asymptotic results with generalized estimating equations for longitudinal data. {\it The Annals of Statistics}, 33(2), 522--541.

\bibitem[{Basu and Santra(2010)}]{basu2010joint}
Basu, S. \& Santra, S. (2010) A joint model for incomplete data in crossover trials. {\it Journal of Statistical Planning and Inference}, 140(10), 2839--2845.

\bibitem[{Cruz et~al.(2023{\natexlab{a}})Cruz, Melo, and Martinez}]{cruz2023correlation}
Cruz, N., Melo, O. \& Martinez, C. (2023) A correlation structure for the analysis of gaussian and non-gaussian responses in crossover experimental designs with repeated measures. {\it Statistical Papers},, 1--28.doi:10.1007/s00362-022-01391-z.

\bibitem[{Cruz et~al.(2023{\natexlab{b}})Cruz, Melo, and Martinez}]{CruzSemi2023}
Cruz, N.A., Melo, O.O. \& Martinez, C.A. (2023) Semiparametric generalized estimating equations for repeated measurements in cross-over designs. {\it Statistical Methods in Medical Research}, 32(5), 1033--1050. doi:10.1177/09622802231158736.

\bibitem[{Curtin(2017)}]{curtin2017meta}
Curtin, F. (2017) Meta-analysis combining parallel and crossover trials using generalised estimating equation method. {\it Research Synthesis Methods}, 8(3), 312--320.

\bibitem[{Davis(2002)}]{Davis}
Davis, C.S. (2002) {\it Statistical Methods for the Analysis of Repeated Measurements}. San Diego: Springer.

\bibitem[{Fleiss(1989)}]{fleiss1989}
Fleiss, J.L. (1989) A critique of recent research on the two-treatment crossover design. {\it Controlled clinical trials}, 10(3), 237--243.

\bibitem[{Grayling et~al.(2018)Grayling, Mander, and Wason}]{grayling2018blinded}
Grayling, M.J., Mander, A.P. \& Wason, J.M. (2018) Blinded and unblinded sample size reestimation in crossover trials balanced for period. {\it Biometrical Journal}, 60(5), 917--933.

\bibitem[{Hao et~al.(2015)Hao, von Rosen, and von Rosen}]{hao2015explicit}
Hao, C., von Rosen, D. \& von Rosen, T. (2015) Explicit influence analysis in two-treatment balanced crossover models. {\it Mathematical Methods of Statistics}, 24(1), 16--36.

\bibitem[{Inan et~al.(2017)Inan, Zhou, and Wang}]{PGEE}
Inan, G., Zhou, J. \& Wang, L. (2017) {\it PGEE: Penalized Generalized Estimating Equations in High-Dimension}. CRAN, r package version 1.5.
\newline\urlprefix\url{https://CRAN.R-project.org/package=PGEE}

\bibitem[{Jaman et~al.(2025)Jaman, Wang, Ertefaie, Bally, L{\'e}vesque, Platt, and Schnitzer}]{jaman2025penalized}
Jaman, A., Wang, G., Ertefaie, A., Bally, M., L{\'e}vesque, R., Platt, R.W. et~al. (2025) Penalized g-estimation for effect modifier selection in a structural nested mean model for repeated outcomes. {\it Biometrics}, 81(1), ujae165.

\bibitem[{Jankar and Mandal(2021)}]{jankaroptimal}
Jankar, J. \& Mandal, A. (2021) Optimal crossover designs for generalized linear models: An application to work environment experiment. {\it Statistics and Applications}, 19(1), 319--336.

\bibitem[{Jankar et~al.(2020)Jankar, Mandal, and Yang}]{jankar2020optimal}
Jankar, J., Mandal, A. \& Yang, J. (2020) Optimal crossover designs for generalized linear models. {\it Journal of Statistical Theory and Practice}, 14(2), 1--27.

\bibitem[{Jones and Kenward(2015)}]{ken15}
Jones, B. \& Kenward, M.G. (2015) {\it Design and Analysis of Cross-Over Trials Third Edition}. Boca Raton: Chapman \& Hall/CRC.

\bibitem[{Josephy et~al.(2015)Josephy, Vansteelandt, Vanderhasselt, and Loeys}]{josephy2015within}
Josephy, H., Vansteelandt, S., Vanderhasselt, M.A. \& Loeys, T. (2015) Within-subject mediation analysis in ab/ba crossover designs. {\it The international journal of biostatistics}, 11(1), 1--22.

\bibitem[{Kitchenham et~al.(2018)Kitchenham, Madeyski, and Curtin}]{kitchenham2018corrections}
Kitchenham, B., Madeyski, L. \& Curtin, F. (2018) Corrections to effect size variances for continuous outcomes of cross-over clinical trials. {\it Statistics in medicine}, 37(2), 320--323.

\bibitem[{Lee et~al.(2022)Lee, Yang, and Wang}]{lee2022doubly}
Lee, D., Yang, S. \& Wang, X. (2022) Doubly robust estimators for generalizing treatment effects on survival outcomes from randomized controlled trials to a target population. {\it Journal of causal inference}, 10(1), 415--440.

\bibitem[{Li et~al.(2018)Li, Forbes, Turner, and Preisser}]{li2018power}
Li, F., Forbes, A.B., Turner, E.L. \& Preisser, J.S. (2018) Power and sample size requirements for gee analyses of cluster randomized crossover trials. {\it Statistics in Medicine},.

\bibitem[{Li et~al.(2020)Li, Gao, and Xu}]{LasoGEE}
Li, Y., Gao, X. \& Xu, W. (2020) {\it LassoGEE: High-Dimensional Lasso Generalized Estimating Equations}. CRAN, r package version 1.0.
\newline\urlprefix\url{https://CRAN.R-project.org/package=LassoGEE}

\bibitem[{Liang and Zeger(1986)}]{liang1986longitudinal}
Liang, K.Y. \& Zeger, S.L. (1986) Longitudinal data analysis using generalized linear models. {\it Biometrika}, 73(1), 13--22.

\bibitem[{Lu(2024)}]{lu2024penalized}
Lu, M. (2024) Penalized estimation of panel count data using generalized estimating equation. {\it Electronic Journal of Statistics}, 18(1), 1603--1642.

\bibitem[{Lui(2015)}]{lui2015test}
Lui, K.J. (2015) Test equality between three treatments under an incomplete block crossover design. {\it Journal of biopharmaceutical statistics}, 25(4), 795--811.

\bibitem[{Madeyski and Kitchenham(2018)}]{madeyski2018effect}
Madeyski, L. \& Kitchenham, B. (2018) Effect sizes and their variance for ab/ba crossover design studies. {\it Empirical Software Engineering}, 23(4), 1982--2017.

\bibitem[{McCulloch and Searle(2004)}]{mcculloch2004generalized}
McCulloch, C.E. \& Searle, S.R. (2004) {\it Generalized, linear, and mixed models}. : John Wiley \& Sons.

\bibitem[{Oh et~al.(2003)Oh, Ko, and Oh}]{oh2003bayesian}
Oh, H.S., Ko, S.g. \& Oh, M.S. (2003) A bayesian approach to assessing population bioequivalence in a 2 2 2 crossover design. {\it Journal of Applied Statistics}, 30(8), 881--891.

\bibitem[{Pitchforth et~al.(2020)Pitchforth, Nelson-White, van~den Helder, and Oosting}]{pitchforth2020work}
Pitchforth, J., Nelson-White, E., van~den Helder, M. \& Oosting, W. (2020) The work environment pilot: An experiment to determine the optimal office design for a technology company. {\it PloS one}, 15(5), e0232943.

\bibitem[{Rosenkranz(2015)}]{rosenkranz2015analysis}
Rosenkranz, G.K. (2015) Analysis of cross-over studies with missing data. {\it Statistical methods in medical research}, 24(4), 420--433.

\bibitem[{Senn(1992)}]{senn1992simple}
Senn, S. (1992) Is the ‘simple carry-over’model useful? {\it Statistics in Medicine}, 11(6), 715--726.

\bibitem[{Vegas et~al.(2016)Vegas, Apa, and Juristo}]{vegas2016crossover}
Vegas, S., Apa, C. \& Juristo, N. (2016) Crossover designs in software engineering experiments: Benefits and perils. {\it IEEE Transactions on Software Engineering}, 42(2), 120--135.

\bibitem[{Wang et~al.(2012)Wang, Zhou, and Qu}]{wang2012penalized}
Wang, L., Zhou, J. \& Qu, A. (2012) Penalized generalized estimating equations for high-dimensional longitudinal data analysis. {\it Biometrics}, 68(2), 353--360.

\bibitem[{Xie and Yang(2003)}]{xie2003asymptotics}
Xie, M. \& Yang, Y. (2003) Asymptotics for generalized estimating equations with large cluster sizes. {\it The Annals of Statistics}, 31(1), 310--347.

\end{thebibliography}
\end{document}